\documentclass[a4paper,fleqn,usenatbib]{mnras}

\usepackage[T1]{fontenc}
\usepackage{ae,aecompl}

\usepackage{graphicx}	
\usepackage{amsmath}	
\usepackage{amssymb}	
\usepackage{subcaption}
\usepackage{amsmath,amsfonts,amssymb}
\usepackage{graphicx}
\usepackage{setspace}
\usepackage{tocloft}

\usepackage{longtable}
\usepackage{float}
\usepackage{subcaption}
\usepackage{booktabs}
 \usepackage{multirow}

\title{Multi-parameter study for a new Ground-Based telescope in Egypt}

\author[M. Darwish et al.]{
Mohamed S. Darwish $^{1,2}$,\thanks{E-mail: darwish.msk@gmail.com}
Hazem Badreldin$^{3,4}$,
Nasser M.  Ahmed$^{1}$,
Mostafa Morsy$^{5}$,
\newauthor
E. E. Kohil$^{6}$,
Hany M. Hassan$^{3,4}$,
I. Helmy$^{1}$,
Ahmed shokry$^{1,2}$,
M. A. Hassan$^{6}$,
\newauthor
S. M. Saad$^{1,2}$,
G. M. Hamed$^{1}$,
Z. F. Ghatass$^{6}$ and
S. A. Ata$^{1}$.
\newauthor
\\
$^{1}$Astronomy Department, National Research Institute of Astronomy and Geophysics (NRIAG), 11421 Helwan, Cairo, Egypt.\\
$^{2}$Kottamia Center of Scientific Excellence in Astronomy and Space Science (KCScE, STDF, ASRT), 11421 Helwan, Cairo, Egypt.\\
$^{3}$Seismology Department, National Research Institute of Astronomy and Geophysics (NRIAG), 11421 Helwan, Cairo, Egypt.\\
$^{4}$North African Group for Earthquakes and Tsunami Studies (NAGET), ICTP, Italy.\\
$^{5}$Astronomy and Meteorology Department, Faculty of Science, Al-Azhar University, 11884 Cairo, Egypt.\\
$^{6}$Department of Environmental Studies, Institute of Graduate Studies and Research, Alexandria University, 832 Shatby, Alexandria, Egypt.\\
}
\date{Accepted XXX. Received YYY; in original form ZZZ}

\pubyear{2023}

\begin{document}
\label{firstpage}
\pagerange{\pageref{firstpage}--\pageref{lastpage}}
\maketitle

%
\begin{abstract}
{A multi-parameter analysis was conducted to evaluate the impact of meteorological parameters, night sky brightness and seismic hazard on proposed sites for the new optical/infrared Egyptian astronomical telescope. The ERA5 reanalysis data set is used to get the following meteorological parameters: Total cloud coverage fraction, precipitable water vapor, relative humidity, wind speed \& direction and Air temperature. To estimate the aerosol optical depth we used the Modern-Era Retrospective analysis for Research and Applications version 2 (MERRA-2). Light pollution over the candidate sites was measured from Visible Infrared Imaging Radiometer Suite (VIIRS) Day Night Band (DNB). The seismic input in terms of maximum acceleration and response spectra were computed using a physics-based ground motion approach to assess the seismic hazards and consequently the designation of seismic resistant structure for the proposed sites to be able to assess the seismic hazards for the candidate sites. Of the seven nominated sites, two sites are found to have the best measurements and might be considered future sites for the new Egyptian Astronomical telescope. The first site is located in the south of the Sinai peninsula, while the second one is located in the Red Sea mountains region.}
\end{abstract}


%
\section{Introduction}\label{sec:intro}

{The geographical location of Egypt is one of the most important reasons for building astronomical telescopes that would bridge the gap in observations between the northern and southern hemisphere. Kottamia Astronomical Observatory (KAO) (at 29$^{\circ}$ 55$'$ 35$"$.24 N, 31$^{\circ}$ 49$'$ 45$"$.85 W), which hosted a 1.88 meter optical telescope and operated by the National Research Institute of Astronomy and Geophysics(NRIAG), is considered the largest optical telescope in the Middle East and North Africa region (MENA), so far. The unique site of KAO enables the Egyptian's researchers to carry out galactic and extra-galactic research, as well as stellar variability and stellar evolution\citep[]{2016NewA...47...24S,2017NewA...50...37D,2017NewA...50...12D}, in addition to discovering a list of variable stars} \citep[e.g.][]{2017NewA...53...35D,2017NewA...55...27S,ABDELSABOUR2024102100}.
\begin{figure*}
 \begin{center}
  \includegraphics[angle=0,width=18cm]{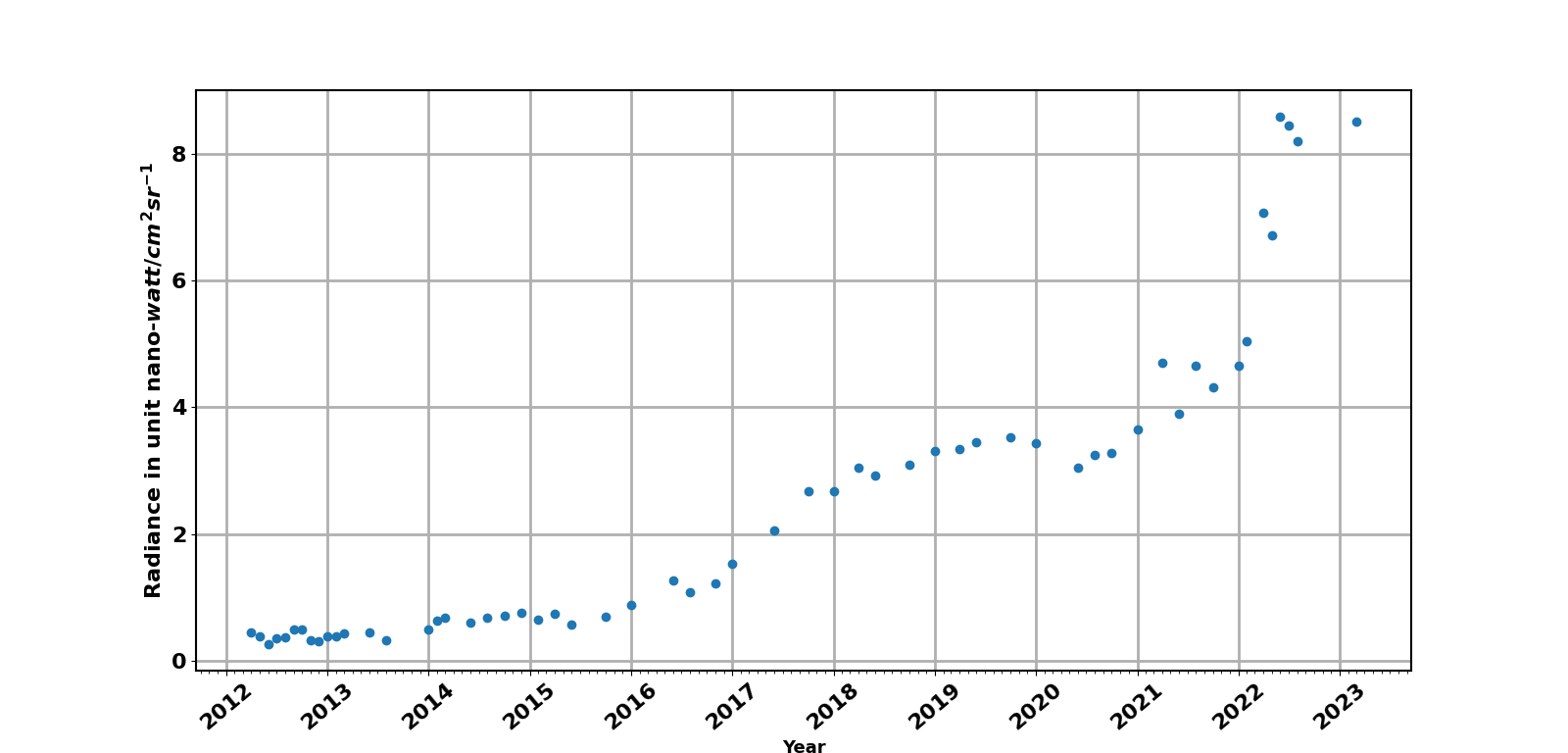}
 \end{center}
  \caption{\label{fig:KAO_LP.png} The average Night sky brightness above KAO measured in nano W/cm\textsuperscript{2}sr\textsuperscript{-1} during the period from 2012 to 2023}.
  \end{figure*} %

{More recently, and owing to the rapid increase of the artificial light at night (ALAN) pollution coming from the New Administrative Capital, the faintest observable magnitude at the KAO site has shifted towards a brighter magnitude. Consequently, objects fainter than 18 mag are no longer observed. Figure \ref{fig:KAO_LP.png} depicts the evolution of Sky brightens above KAO using the Visible Infrared Imaging Radiometer Suite (VIIRS)-Day Night Band (DNB) during the period of 2012 to 2022, where DNB is in visible band (0.5 - 0.9 $\mu$m) (\cite{Nurbandi_2016}. Therefore, a new contemporary optical/IR telescope (> 4 m) is needed.}

{The first step is to think about the quality of the astronomical site, which consequently leads to the best performances from ground-based optical telescopes. This quality can be characterized by several conditions. One of these conditions is the atmospheric parameters including Air temperature (AT), wind speed (WS), wind direction (WD), relative humidity (RH), precipitable water vapour (PWV), total cloud coverage (TCC) and aerosol optical depth (AOD), parameters that play an important role in the quality of astronomical observations from ground-based sites \citep[e.g.][]{1983ESOC...17..217A,2009MNRAS.399..783L,2009otam.conf..256V}. Secondly, the sky brightness and the light pollution which is also an issue affecting the quality of the astronomical observations and their limiting magnitude}.

{The night sky brightness as seen from the ground is mainly due to natural or artificial sources. These natural sources are the extra-terrestrial (e.g. unresolved stars/galaxies, diffuse galactic background and zodiacal light) and atmospheric phenomena (e.g. auroral activity and air-glow), while artificial components are the so called artificial light which scattered by the troposphere and caused by human activity \citep[e.g.][]{1964SSRv....3..512R, 1998ASPC..139...17L,2003A&A...400.1183P, 2004PASP..116..762T,masana2021multiband, barentine2022night}. Last but not least, site accessibility and seismic hazard parameters are also important considerations. The seismic hazard in particular, can play a crucial role in construction, operational as well as survival conditions of telescope both in structure and cost \citep[e.g][]{2008SPIE.7012E..4JT, 2010MNRAS.407.1361E,usuda2014preliminary,sugimoto2022seismic}.

{Considering the MENA region, a few studies have been done in order to search for the best sites for an optical telescope \citep[e.g][]{abdelaziz2017search}. \citet{abdelaziz2017search} focused only on some meteorological parameters namely, AT, Barometric Pressure (BP), RH, WD, WS, AOD and PWV. They concluded that, as for Egypt the best site to set an optical telescope is located at the Egyptian western desert. More recently, \citet{2020MNRAS.493.1204A} presented some meteorological parameters to select not only MENA but global astronomical sites. We will talk about this work in details in Sec. \ref{sec:sites}.}

{In the present work, we aim to investigate the meteorological conditions, night sky brightness and seismic hazard assessment for a list of candidate sites proposed for the new Egyptian Large Optical Telescope.}
{In Section \ref{sec:sites}, the selection criteria for the candidate sites are described. The meteorological conditions including their parameters as well as the Light Pollution for each site are given in section \ref{sec:meteoro}. Section \ref{sec:s_hazard} deals with the seismic hazard assessment. Finally, our conclusion is presented in section \ref{sec:conc_reco}.}


\section{Candidate sites and selection criteria}\label{sec:sites}
 {A fundamental parameter to search a ground-based telescope site is to look at the region's spatial information or topography which is usually presented by Digital Elevation Models (DEM) maps. In order to generate the DEM maps for Egypt, the NASA's Shuttle Radar Topography Mission (SRTM V3) with resolution around 30 m is used (See, \cite{farr2007shuttle}). The high resolution provided by SRTM V3 enables us to locate very good curvatures and plateaus that might be a candidate site for a new ground-based telescope. Figure \ref{fig:DEM_all.png} shows the full map of Egypt. preliminary selection criteria including, Elevation (> 1000 m), site accessibility, distance from the city lights ($\geq$ 50 km) from the nearest city and Night Sky Brightness (NSB) to be fainter than 21.85 $mag./arcsec^2$}.
 {Following these criteria, a number of sites are selected, listed and displayed in Table \ref{tab:candidate_sites} and Figure \ref{fig:DEM_all.png}, respectively.}

\begin{table}
\caption{ List of the candidate sites ID, name, longitude, latitude and elevation.}
\label{tab:candidate_sites}
\begin{center}
\renewcommand{\arraystretch}{1.1}
\begin{tabular}{|c|c|c||c|c|}
\hline
 Site No. & N Latitude (deg)& E Longitude (deg)& Elevation(m)\\
\hline

1 & 28.847995 & 34.096922 & 1583 \\
2 & 28.880113 & 33.891536 & 1612\\
3 & 29.045706 & 33.910842 & 1626\\
4 & 27.470417 & 33.011806 &1381\\
5 & 26.978606 & 33.487795 & 2100\\
6 & 27.027083 & 33.28625  & 1631 \\
7 & 26.966806 & 33.332917 & 1531\\
8 & 22.7185   & 34.696    &	1315\\
\hline
\end{tabular}
      \end{center}
      \end{table} %

\begin{figure*}
  \includegraphics[angle=0,width=15cm]{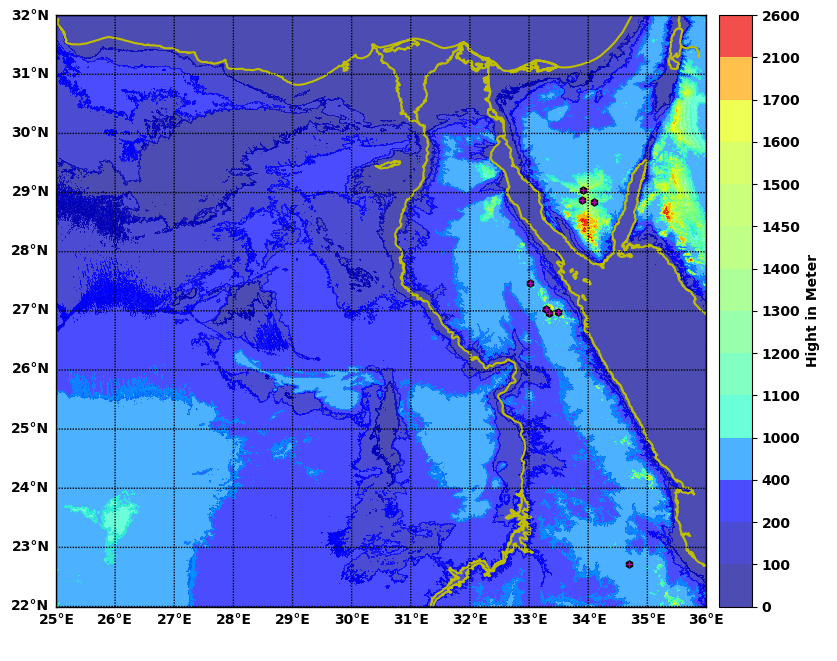}
  \caption{\label{fig:DEM_all.png}Topography of Egypt along with the Candidate Mountains (black dots), the elevation through the map is colour coded.}
  \end{figure*} %

{In order to confirm those sites as good candidates, we first compared them with the high-resolution satellite global data published by \citet{2020MNRAS.493.1204A}. They introduced an index to evaluate the site's suitability named "Suitability Index for Astronomical Sites" (SIAS), with A, B, C and D series. Of this data, only data above 1000 m for Egypt are considered.}
\begin{figure*}
  \includegraphics[angle=0,width=14.5cm]{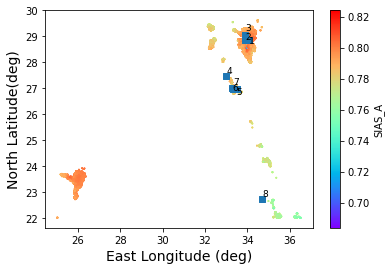}
  \caption{\label{fig:SIAS_A}Series A of \citet{2020MNRAS.493.1204A} Suitability Index for Astronomical Sites (SIAS), the 8 candidate sites are marked with blue squares, where the SIAS index through the map is colour coded.}
  \end{figure*} %
{As shown in Figure \ref{fig:SIAS_A}, most of the candidate sites are in a good agreement with \citet{2020MNRAS.493.1204A} criteria except site No. 8 (see Table \ref{tab:candidate_sites}) which consequently will not be further taken into account.}
\section{sites meteorological conditions}\label{sec:meteoro}
{In this section, we focus on some important meteorological parameters which would help in the site evaluation process. These parameters are, near surface (2 m) Air Temperature (T,$^{\circ}$C) and Relative humidity (RH, \%), Wind speed (WS, ms$^{-1}$) and direction (WD, degree) at 10 m, Precipitable water vapor (PWV, mm) and Total Cloud Cover (TCC, fraction from 0 to 1).}

{These parameters were obtained as monthly averaged data from ERA5 reanalysis dataset \citep{81046} at 0.25$^{\circ}$ x 0.25$^{\circ}$ grid spatial resolution for the climate period of 40 years (1979-2019). The data were extracted at the candidate sites using bilinear interpolation between the nearest 4 grid points. ERA5 is the fifth and latest generation of European Centre for Medium-Range Weather Forecasts (ECMWF) global atmospheric (climate and weather) reanalysis data set, which combines vast amounts of historical observations into global estimates using advanced modelling and data assimilation systems. For more details on ERA5 see \url{https://www.ecmwf.int/en/forecasts/datasets/reanalysis-datasets/era5}.}

{Owing to the ERA5 spatial resolution (0.25$^{\circ}$ x 0.25$^{\circ}$) sites No. 6 and 7 are taken as one site labeled with 6\&7 ID, because of their similarity in interpolated geographical location which give the same parameter values. The local seasons are defined by grouping months as follows, \textit{winter}: December, January and February; \textit{spring}: March, April and May; \textit{summer}: June, July and August; \textit{autumn}: September, October and November.\\
For further statistical analysis and interpretation of daytime and nighttime changes for selected meteorological variables such as Air Temperature and Wind speed, the monthly averaged reanalysis by hour of day from ERA5 dataset is used. Day and night time separation for meteorological analysis was determined by the times of sunset and sunrise; however, for clear nights estimation (see., \ref{cl_nigh}), the nighttime is better defined by nautical twilight.
In addition, the maximum, minimum, mean, standard deviation, and the different percentage (5\%, 25\%, 50\%, 75\%, and 95\%) for each variable are computed to determine its variation (highest and lowest) during the daytime and nighttime at the different sites as demonstrated in Table \ref{tab:AT_hour} and \ref{tab:WS_hour}.}
\subsection{Air Temperature}%
{Air temperature is an atmospheric parameter which directly affects the telescope's detectors (e.g. charge coupled device(CCD) and mirror) \citep{1979MNRAS.188..249L,2015PASP..127.1292Z} .
Where an increase or decrease in the ambient air temperature by an amount greater than 2 or less than -2 $^{\circ}$C than the telescope mirror temperature degrades seeing and affects the imaging performance of the telescope \citep{1979MNRAS.188..249L,2003SPIE.5179..270V,2004SPIE.5497..497D,2011NewA...16..328B}.
Moreover, the large scale variations of AT lead to pressure gradients and winds which play a significant role in promoting atmospheric turbulence that affects the operation of the telescope and leads to bad seeing \citep[e.g.][]{1979MNRAS.188..249L, 2012ARA&A..50..305D}.}
{In addition to the optical telescope operation, AT is expected to have a serious influence on Radio telescopes leading to thermal deformation of its mechanical structures \citep{2019PASP..131d5001O}.}
Thus, analyzing the temperature at the chosen observatory locations is very important to interpret and understand the temperature gradient and variation.\\
Table \ref{tab:AT_hour} clearly highlights that there is no significant variation in the AT at the nighttime in all the candidate sites. The minimum AT value recorded is found to be in the range from 0.7 to 4 $^{\circ}$C, while for 95\% of the nights the AT ranges between 24 and 30 $^{\circ}$C. Compared with VLT telescope at Paranal, these values are aligning with the safe operation condition.

\begin{table*}
\tiny
\centering
\setlength{\tabcolsep}{3pt}
\caption{Daytime and Nighttime changes of AT for each candidate site, ERA5 monthly averaged reanalysis by hour of day is used.}\label{tab:AT_hour}
\begin{tabular}{|c|*{24}{c}|*{24}{c}|}
\hline
&\textbf{Day Time} & \multicolumn{23}{c}{\textbf{ Night Time}}&  \multicolumn{23}{c}{} \\
\textbf{Time} & 07 & 08 & 09 & 10 & 11 & 12 & 13 & 14 & 15 & 16 & 17 & 18 & 19 & 20 & 21 & 22 & 23 & 00 & 01 & 02 & 03 & 04 & 05 & 06 \\
\hline
Site-1 & 15.36 & 16.88 & 17.85 & 21.25 & 21.98 & 23.69 & 23.76 & 23.69 & 23.47 & 22.50 & 21.43 & 20.95 & 18.52 & 17.80 & 17.47 & 15.90 & 15.31 & 13.80 & 13.66 & 13.50 & 13.47 & 12.71 & 12.89 & 13.10 \\
Site-2 & 14.92 & 16.55 & 17.58 & 21.02 & 21.74 & 23.43 & 23.43 & 23.28 & 23.02 & 22.05 & 21.00 & 20.58 & 18.21 & 17.51 & 17.15 & 15.54 & 14.89 & 13.35 & 13.18 & 13.01 & 12.96 & 12.20 & 12.35 & 12.60 \\
Site-3 & 14.45 & 16.04 & 17.12 & 20.64 & 21.46 & 23.21 & 23.26 & 23.15 & 22.92 & 21.95 & 20.87 & 20.42 & 18.02 & 17.28 & 16.88 & 15.23 & 14.53 & 12.95 & 12.76 & 12.58 & 12.55 & 11.83 & 12.01 & 12.22 \\
Site-4 & 18.89 & 20.19 & 21.27 & 25.08 & 26.00 & 28.12 & 28.37 & 28.52 & 28.53 & 27.86 & 26.94 & 26.36 & 23.62 & 22.90 & 22.58 & 21.02 & 20.42 & 18.75 & 18.46 & 18.14 & 17.88 & 16.77 & 16.65 & 16.72 \\
Site-5 & 20.30 & 21.60 & 22.39 & 25.74 & 26.46 & 28.33 & 28.51 & 28.57 & 28.46 & 27.65 & 26.76 & 26.31 & 23.87 & 23.27 & 22.94 & 21.41 & 20.85 & 19.26 & 19.07 & 18.84 & 18.68 & 17.67 & 17.64 & 17.83 \\
Site-6\&7 & 19.83 & 21.11 & 22.03 & 25.51 & 26.29 & 28.26 & 28.50 & 28.64 & 28.62 & 27.93 & 27.12 & 26.69 & 24.14 & 23.47 & 23.08 & 21.46 & 20.84 & 19.19 & 18.96 & 18.71 & 18.53 & 17.51 & 17.48 & 17.57 \\
\hline
\textbf{Statistics} & \textbf{Max} & \textbf{Min} & \textbf{Mean} & \textbf{SD} & \textbf{5\%} & \textbf{25\%} & \textbf{50\%} & \textbf{75\%} & \textbf{95\%} & & \textbf{Max} & \textbf{Min} & \textbf{Mean} & \textbf{SD} & \textbf{5\%} & \textbf{25\%} & \textbf{50\%} & \textbf{75\%} & \textbf{95\%} \\
\hline
Site-1 & 34.34 & 1.58 & 21.07 & 7.21 & 8.51 & 15.24 & 21.98 & 27.32 & 31.10 & & 29.04 & 0.95 & 14.84 & 6.39 & 4.71 & 9.28 & 15.67 & 19.91 & 24.60 \\
Site-2 & 34.13 & 1.21 & 20.72 & 7.19 & 8.09 & 14.89 & 21.68 & 26.98 & 30.64 & & 28.84 & 0.71 & 14.41 & 6.32 & 4.43 & 8.94 & 15.19 & 19.31 & 24.30 \\
Site-3 & 33.84 & 1.36 & 20.46 & 7.12 & 7.90 & 14.80 & 21.29 & 26.67 & 30.32 & & 28.51 & 0.93 & 14.07 & 6.21 & 4.29 & 8.67 & 14.79 & 18.75 & 24.05 \\
Site-4 & 38.29 & 4.23 & 25.51 & 7.48 & 11.85 & 19.78 & 26.41 & 32.11 & 35.83 & & 34.22 & 3.86 & 19.49 & 6.89 & 8.43 & 13.62 & 20.08 & 24.91 & 30.14 \\
Site-5 & 37.84 & 4.95 & 25.92 & 7.01 & 13.39 & 20.43 & 26.91 & 32.09 & 35.53 & & 34.03 & 4.43 & 20.11 & 6.74 & 9.24 & 14.31 & 20.73 & 25.56 & 30.29 \\
Site-6\&7 & 38.26 & 4.39 & 25.88 & 7.31 & 12.73 & 20.14 & 26.88 & 32.35 & 35.81 & & 34.55 & 3.96 & 20.08 & 6.88 & 8.94 & 14.20 & 20.80 & 25.48 & 30.63 \\
\hline
\end{tabular}
\end{table*}

 {Figure \ref{fig:Temp_all} illustrates the monthly climate average of the 2 m temperature for the proposed seven sites. Where, sites 1, 2, and 3 have lower temperature than the sites 4, 5 and 6\&7  along the year. Furthermore, for all sites, the minimum temperature is noticed during winter and gradually increases during spring followed by autumn, while the maximum temperature is seen during summer as shown in Table \ref{tab:climate}. In addition, the lowest annual climate average temperature (16.94$^{\circ}$C) is detected at site 3, while the highest one (22.94) is observed at site 5.}\\

\begin{figure}
  \includegraphics[angle=0,width=8.5cm]{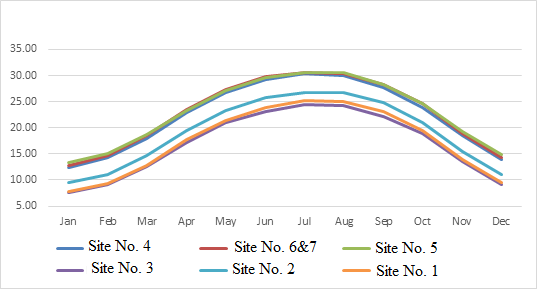}
  \caption{\label{fig:Temp_all}Monthly average of the Air Temperature (AT) during the period from 1979 to 2019 over the candidate sites.}
  \end{figure} %

\subsection{Wind speed and wind direction}
{Wind gusts or strong winds represent a critical hazard for the telescope's instruments and therefore constrain the telescope’s operation \citep{2016PASP..128c5004T}. Figure \ref{wind-1} right side displays the histogram of the WS through all the candidate sites.\\
The higher temporal resolution analysis for WS illustrated in Table \ref{tab:WS_hour} indicates that the nighttime maximum value for the candidate sites is in the range from 3.8 to 5.2 m/s.}

{On the other hand, the maximum monthly average WS values is  found to be 3.86 m/s at site No. 4, while site No. 1 shows the minimum WS measurements with 2.65 m/s.These results (either the hourly or the monthly average) are in all cases lower than the safe operation limits (15 m/s) suggested by \citet{1985VA.....28..449M}. This also agrees with best condition to operate the VLT telescope at Paranal (< 12 m/s). Table \ref{tab:climate} lists the monthly as well as the annual WS average measurements. As one can notice from Table \ref{tab:climate}, the WS values are slightly higher in the summer compared to winter.}\\

\begin{table*}
\tiny
\centering
\setlength{\tabcolsep}{3pt}
\caption{Daytime and Nighttime changes of WS for each candidate site, ERA5 monthly averaged reanalysis by hour of day is used.}\label{tab:WS_hour}
\begin{tabular}{|c|*{24}{c}|*{24}{c}|}
\hline\\
&\textbf{Day Time} & \multicolumn{23}{c}{\textbf{ Night Time}}&  \multicolumn{23}{c}{} \\
\textbf{Time} & 07 & 08 & 09 & 10 & 11 & 12 & 13 & 14 & 15 & 16 & 17 & 18 & 19 & 20 & 21 & 22 & 23 & 00 & 01 & 02 & 03 & 04 & 05 & 06 \\
\hline
Site-1 & 1.11 & 1.55 & 1.86 & 1.91 & 1.90 & 1.96 & 2.14 & 2.36 & 2.61 & 2.84 & 2.95 & 2.85 & 2.64 & 2.41 & 2.09 & 1.79 & 1.54 & 1.31 & 1.25 & 1.20 & 1.16 & 1.11 & 1.05 & 0.97 \\
Site-2 & 1.15 & 1.26 & 1.39 & 1.50 & 1.70 & 2.02 & 2.42 & 2.76 & 3.01 & 3.14 & 3.11 & 2.84 & 2.48 & 2.14 & 1.79 & 1.48 & 1.25 & 1.01 & 0.96 & 0.96 & 0.99 & 1.04 & 1.09 & 1.13 \\
Site-3 & 0.92 & 1.13 & 1.44 & 1.73 & 2.00 & 2.32 & 2.74 & 3.13 & 3.45 & 3.64 & 3.59 & 3.23 & 2.75 & 2.31 & 1.87 & 1.48 & 1.18 & 0.91 & 0.81 & 0.77 & 0.77 & 0.79 & 0.82 & 0.86 \\
Site-4 & 2.21 & 2.80 & 3.36 & 3.48 & 3.32 & 3.10 & 3.14 & 3.31 & 3.56 & 3.80 & 3.90 & 3.75 & 3.43 & 3.21 & 3.07 & 3.01 & 2.95 & 2.74 & 2.67 & 2.60 & 2.53 & 2.44 & 2.31 & 2.15 \\
Site-5 & 1.88 & 2.34 & 2.81 & 2.99 & 2.99 & 2.92 & 2.96 & 3.07 & 3.20 & 3.25 & 3.03 & 2.62 & 2.36 & 2.26 & 2.23 & 2.22 & 2.20 & 2.10 & 2.08 & 2.07 & 2.05 & 2.00 & 1.94 & 1.85 \\
Site-6\&7 & 1.66 & 2.05 & 2.45 & 2.52 & 2.46 & 2.42 & 2.55 & 2.71 & 2.86 & 2.94 & 2.84 & 2.57 & 2.31 & 2.19 & 2.11 & 2.05 & 1.99 & 1.83 & 1.79 & 1.78 & 1.78 & 1.76 & 1.71 & 1.63 \\
\hline
\textbf{Statistics} & \textbf{Max} & \textbf{Min} & \textbf{Mean} & \textbf{SD} & \textbf{5\%} & \textbf{25\%} & \textbf{50\%} & \textbf{75\%} & \textbf{95\%} & & \textbf{Max} & \textbf{Min} & \textbf{Mean} & \textbf{SD} & \textbf{5\%} & \textbf{25\%} & \textbf{50\%} & \textbf{75\%} & \textbf{95\%} \\
\hline
Site-1 & 5.50 & 0.07 & 2.17 & 1.04 & 0.68 & 1.37 & 2.06 & 2.84 & 4.12 & & 4.76 & 0.05 & 1.54 & 0.85 & 0.57 & 0.98 & 1.33 & 1.79 & 3.51 \\
Site-2 & 5.30 & 0.03 & 2.19 & 1.10 & 0.52 & 1.35 & 2.10 & 2.96 & 4.18 & & 4.63 & 0.03 & 1.36 & 0.85 & 0.31 & 0.77 & 1.22 & 1.63 & 3.33 \\
Site-3 & 5.93 & 0.00 & 2.45 & 1.32 & 0.53 & 1.35 & 2.34 & 3.39 & 4.85 & & 5.07 & 0.02 & 1.28 & 0.97 & 0.29 & 0.66 & 0.97 & 1.50 & 3.59 \\
Site-4 & 6.67 & 0.44 & 3.31 & 1.13 & 1.54 & 2.50 & 3.24 & 4.14 & 5.22 & & 5.85 & 0.54 & 2.76 & 0.95 & 1.48 & 2.00 & 2.62 & 3.42 & 4.49 \\
Site-5 & 5.38 & 0.72 & 2.84 & 0.88 & 1.49 & 2.16 & 2.78 & 3.51 & 4.33 & & 3.94 & 0.52 & 2.11 & 0.52 & 1.34 & 1.74 & 2.07 & 2.44 & 3.04 \\
Site-6\&7 & 4.84 & 0.28 & 2.50 & 0.84 & 1.20 & 1.85 & 2.48 & 3.15 & 3.87 & & 3.82 & 0.46 & 1.91 & 0.55 & 1.15 & 1.49 & 1.85 & 2.25 & 2.92 \\
\hline
\end{tabular}
\end{table*}

{Another important wind parameter is its direction. For the astronomical observations the stability of the wind direction is important, since leads to the stability of both airflow and local turbulence \citep{2006PASP..118.1048G}.}

{The annual climate wind rose (Figure \ref{wind-1} left side) illustrates that the dominant wind direction for the most of candidate sites is north to northwest, which could provide a stable local turbulence. While we currently lack the means to present our own graph (due to the lack of seeing measurements), we find support for the link between wind direction and seeing in the work of \cite{atmos14020199}.}
\begin{figure*}
\centering
\begin{subfigure}{.5\textwidth}
  \centering
  \includegraphics[width=8.75cm,height=5cm]{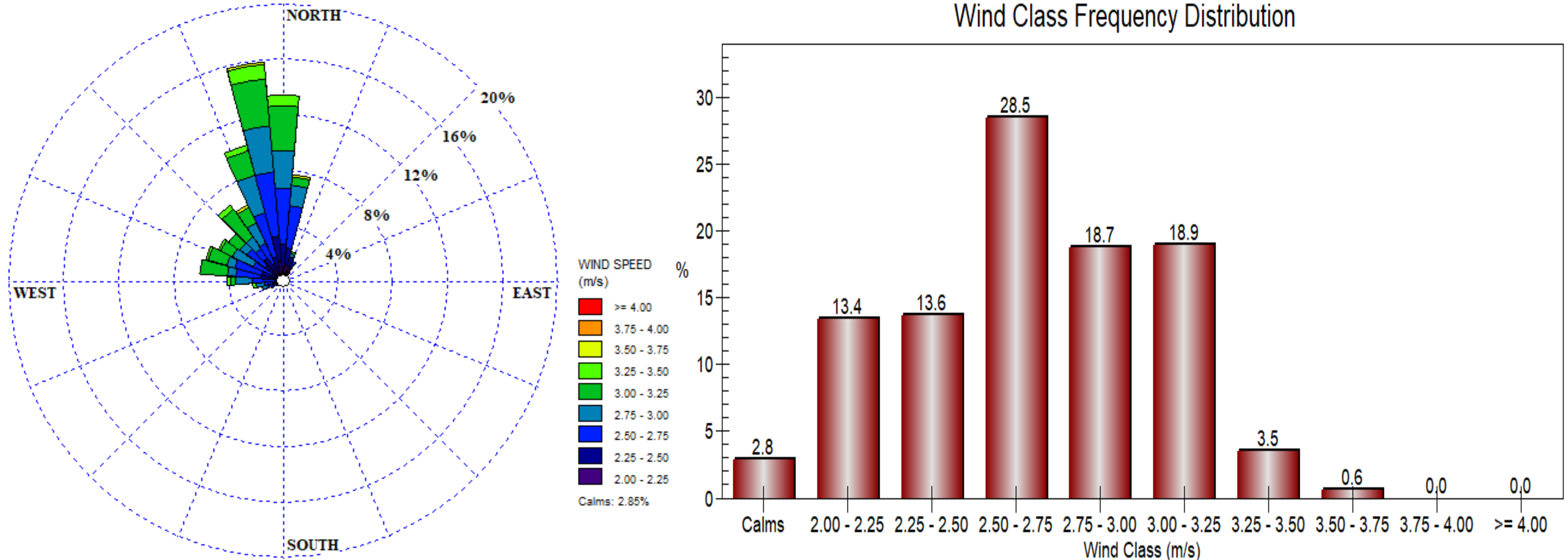}
  \caption{Site No 1}
  \label{wi-1}
\end{subfigure}%
\begin{subfigure}{.5\textwidth}
  \centering
  \includegraphics[width=8.75cm,height=5cm]{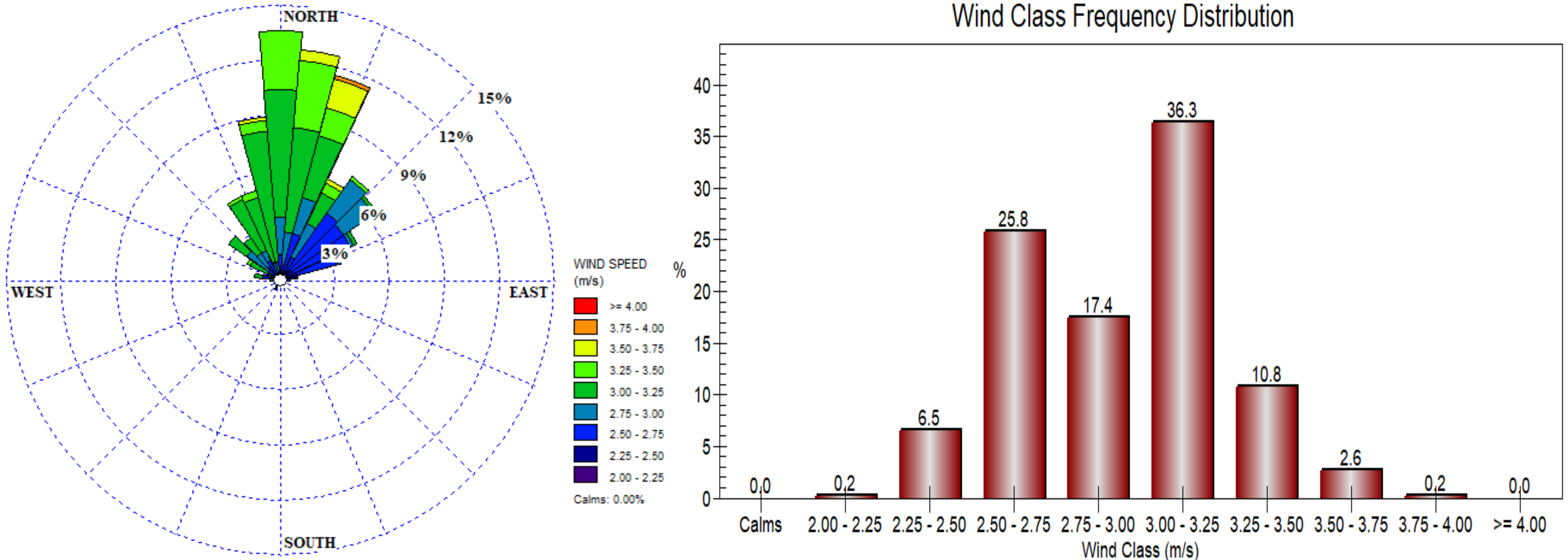}
    \caption{Site No 2}
  \label{wi-2}
\end{subfigure}

\begin{subfigure}{.5\textwidth}
  \centering
  \includegraphics[width=8.75cm,height=5cm]{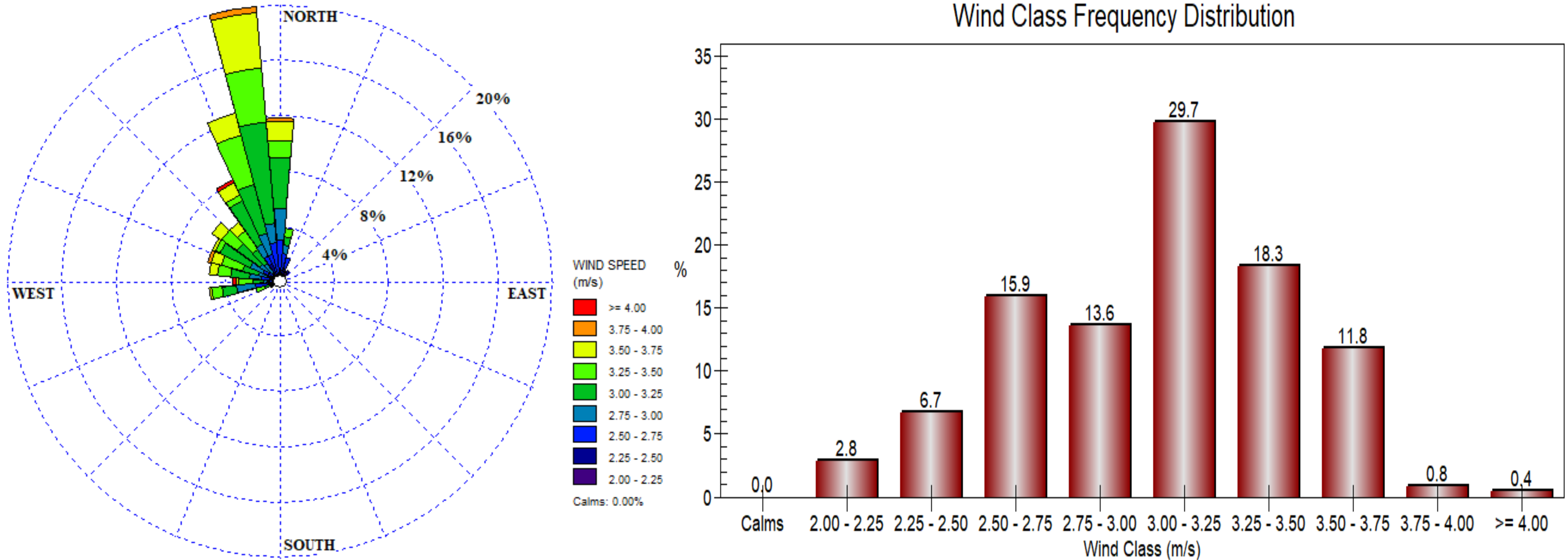}
  \caption{Site No 3}
  \label{wi-3}
\end{subfigure}%
\begin{subfigure}{.5\textwidth}
  \centering
  \includegraphics[width=8.75cm,height=5cm]{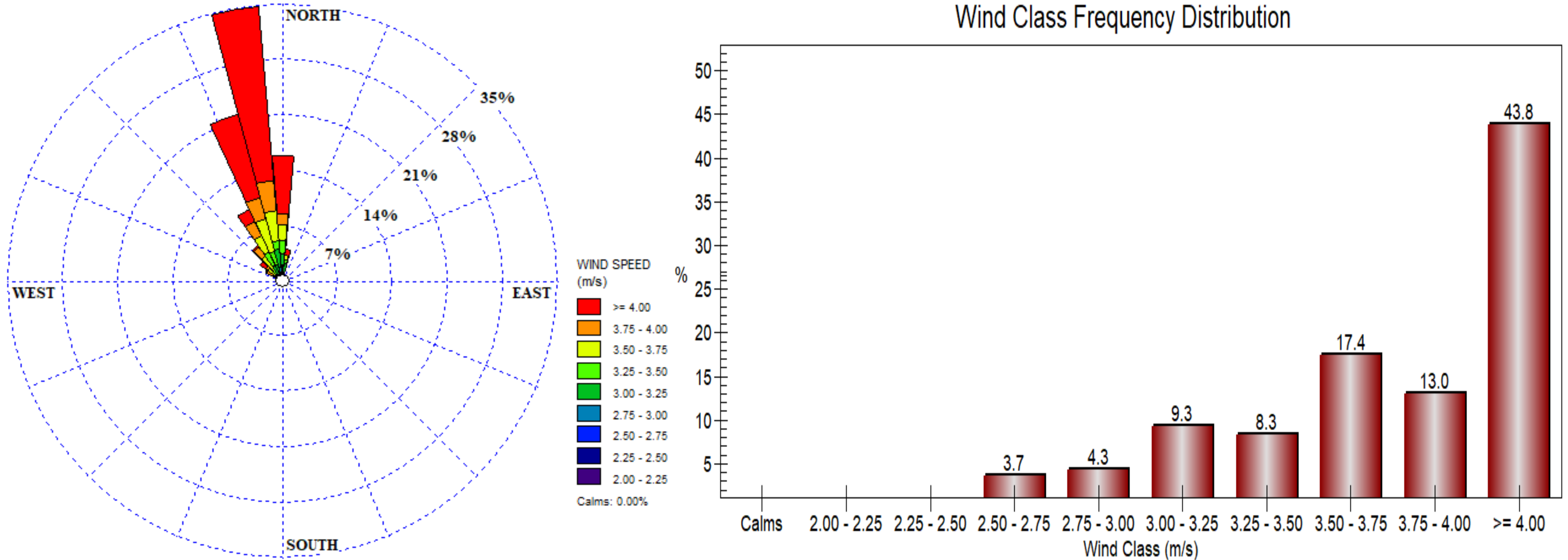}
  \caption{Site No 4}
  \label{wi-4}
\end{subfigure}%

\begin{subfigure}{.5\textwidth}
  \centering
  \includegraphics[width=8.75cm,height=5cm]{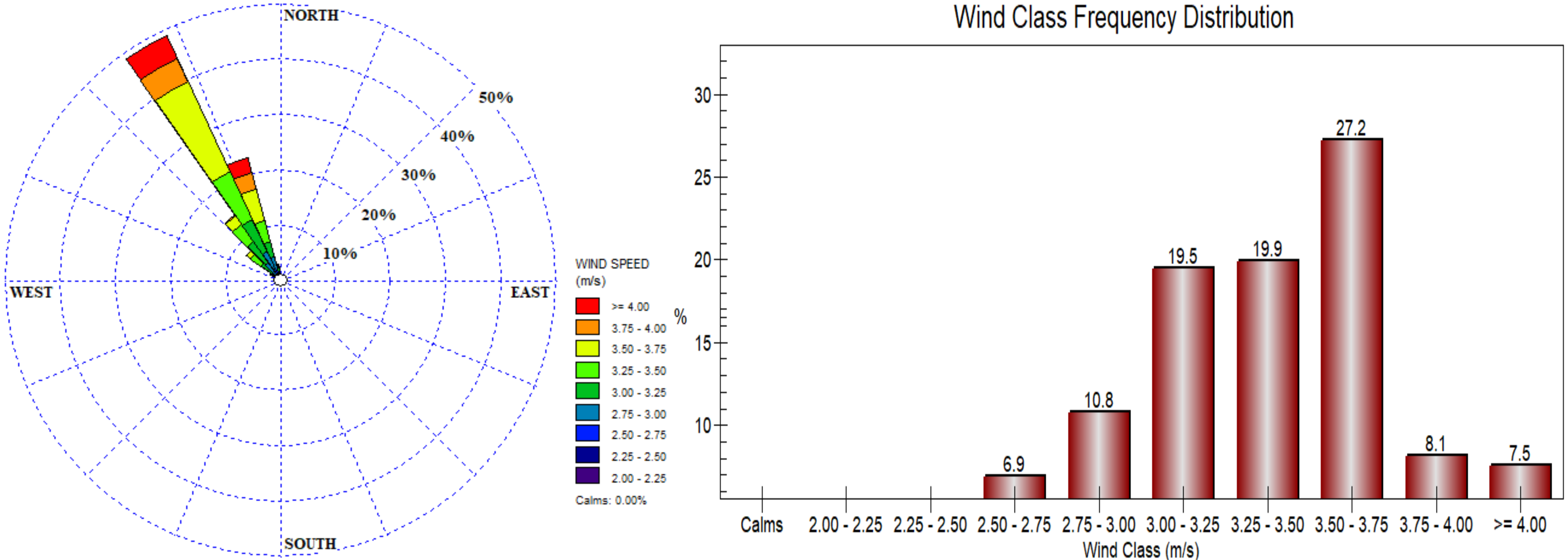}
  \caption{Site No 5}
  \label{wi-5}
\end{subfigure}%
\begin{subfigure}{.5\textwidth}
  \centering
  \includegraphics[width=9cm,height=5cm]{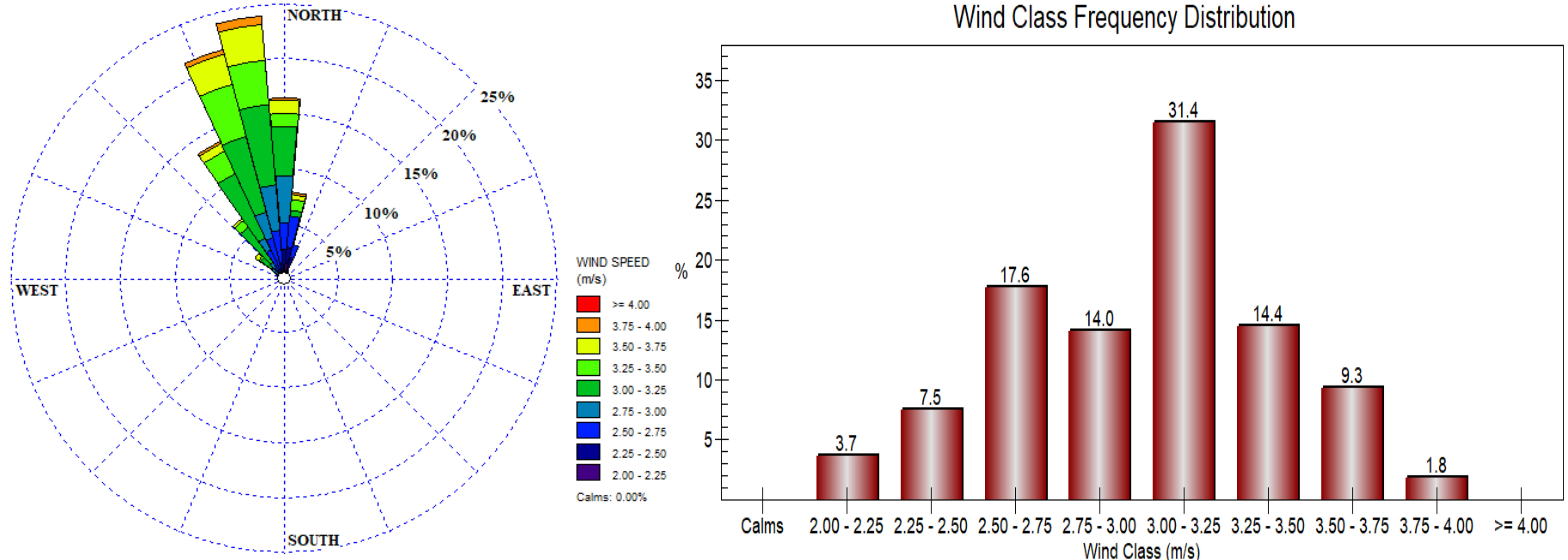}
  \caption{Site No 6\&7}
    \label{wi-67}
\end{subfigure}%
\caption{Wind rose density (left) and Wind speed histograms(right)for each site during the period from 1979 to 2019.}\label{wind-1}
\end{figure*}
%

\subsection{Relative humidity}%
{In order to test the effect of the moisture and water condensation on the telescope's main mirror, instruments and consequently astronomical image quality \citep{2012MNRAS.422.2262R}, one need to measure the Relative humidity (RH). The safety limits of RH which could lead to stopping astronomical observations is $\ge$ 70 \% at Paranal Observatory (Chile).}

{Figure \ref{fig:RH} shows the monthly RH at 2 m measurements for the candidate sites. The four sites (4, 5, 6 and 7) which are located northwest of the Red sea (West of Hurghada) appear to have RH values lower and better than other sites namely 1, 2 and 3. Table \ref{tab:climate} lists the annual as well as the monthly average values, where the summer months seem to have the best RH values. The annual average of RH for all sites indicates that the RH is in general lower than 50\%.}
\begin{figure}
  \includegraphics[angle=0,width=8.5cm]{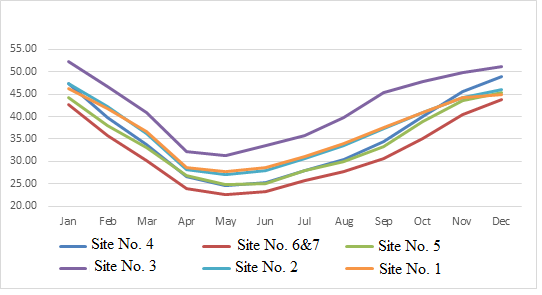}
  \caption{\label{fig:RH}Monthly average of the RH during the period from 1979 to 2019 over the candidate sites.}
  \end{figure} %

\subsection{Precipitable water vapor}
{PWV is defined as the mass of the water for a column of unit size integrated from the surface to the top of the atmosphere. This atmospheric parameter is a crucial factor for telescopes operating at mid-infrared and sub-millimetre regimes \citep{2001JGR...10620101C,2010PASP..122..470O}.}

{Figure \ref{fig:PWV} shows the variation of the monthly averages of the PWV values where the sites No 1, 2 and 3 exhibit lowest measurements especially during winter. The annual values listed in Table \ref{tab:climate} indicate that site No. 1 have the best PWV followed by site No. 3.}

\begin{figure}
  \includegraphics[angle=0,width=8.5cm]{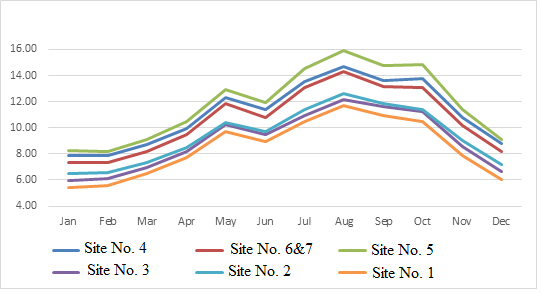}
  \caption{\label{fig:PWV}Monthly average of the PWV during the period from 1979 to 2019 over the candidate sites.}
  \end{figure} %

\subsection{Total Cloud Coverage}%
 {A fundamental and most important parameter for ground-based optical astronomical telescope site selection is the so-called Total cloud coverage (TCC)\citep[e.g.][]{2006Msngr.125...44S,2008MNRAS.391..507V, 2020MNRAS.493.1204A}. The presence of a high percentage of clouds over a certain site would lead to stopping the observations. Monitoring cloud coverage is usually conducted through different methods including, All Sky Camera, satellite monitoring and naked-eye observations \citep{2020RAA....20...83W}.}

{Figure \ref{fig:TCC} shows the monthly TCC over all candidate sites, while Table 2 displays the number of the clear nights for each site during 2019 and the climate period 1979-2019, and Table \ref{tab:climate} lists the monthly and annual climate averages of TCC. It is obvious that the TCC is lower in summer than other seasons followed by autumn and the highest TCC is detected in winter. It is also noted that the red sea sites (4, 5 and 6\&7) have better TCC in comparison with Sinai sites (1, 2 and 3).}

 \begin{figure}
  \includegraphics[angle=0,width=8.5cm]{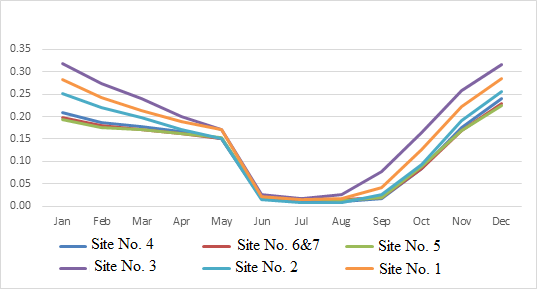}
  \caption{\label{fig:TCC}Monthly average of the TCC during the period from 1979 to 2019 over the candidate sites.}
  \end{figure} %
  \subsection{Clear nights at each site} \label{cl_nigh}
The quality of an optical astronomical observatory is significantly impacted by the number of clear nights per year. The assessment of whether a night is adequately clear can be directly made by evaluating the level of cloud coverage.
To estimate the number of clear nights per year, we followed the definition of clear nights by \citet{2000A&AS..145..293E}, where the clear night should have a cloud coverage less than 25\%. The ERA5 night hourly data (from 6 Pm to 4 Am, local time) during 2019 is used to investigate this parameter.

Table \ref{tab:phot_nights} enumerates number of clear nights per year for each site alongside their respective ratios. Among the potential locations, site No. 2 emerged with the highest count of clear nights at 302, followed by site No. 4. Although the number of clear nights at other sites falls below that of 2 and 4, they too merit consideration as viable candidates for building an observatory due to their cloud coverage. Nonetheless, it is imperative to undertake further measurements and procedures to illustrate the quality of photometric observations at each site, as detailed in \cite{2014SPIE.9149E..0MK}.

 \subsection{Aerosol Optical Depth (AOD)}
{One of the important parameters that the astronomical sites are characterised with is the atmospheric extinction, which directly affects the sky transparency and consequently the quality of the astronomical observations. Such extinction is mainly caused by either precipitable water vapor or aerosols.}

{Atmospheric aerosols comprise a wide range of particle types with different compositions, sizes, shapes, and properties. Aerosols are ubiquitous in air and are often observable as dust, smoke, and haze. Their sizes occupy a wide range covering from 10$^{-2}$ $\mu$m to about 10$^{2}$ $\mu$m, where the most effective size in attenuating sunlight is in the range from 0.1 to 1 $\mu$m \citep{ranjan2007study}. The amount of aerosol in the atmosphere, is usually quantified by mass concentration or by an optical measure; aerosol optical depth (AOD). AOD is affected by various factors such as aerosol sources, weather conditions, landscape, and regional differences.}\\

\begin{table*}
\begin{center}
\caption{List of the climatological parameters both monthly and annual average over the candidate sites.}
\label{tab:climate}
\resizebox{.65\textwidth}{!}{\begin{tabular}{llllllll}
\hline
\multirow{14}{*}{T (ºC)}         & Month  & Site No. 1 & Site No. 2 & Site No. 3 & Site No. 4 & Site No. 5 & Site No. 6\&7 \\
\hline
                                  & Jan    & 7.77       & 9.47       & 7.63       & 12.44      & 13.35      & 12.83         \\
                                  & Feb    & 9.28       & 11.02      & 9.12       & 14.27      & 15.05      & 14.67         \\
                                  & Mar    & 12.82      & 14.61      & 12.55      & 18.02      & 18.62      & 18.53         \\
                                  & Apr    & 17.69      & 19.39      & 17.25      & 22.84      & 23.27      & 23.47         \\
                                  & May    & 21.47      & 23.23      & 20.96      & 26.83      & 27.19      & 27.4          \\
                                  & Jun    & 23.87      & 25.73      & 23.16      & 29.29      & 29.69      & 29.76         \\
                                  & Jul    & 25.14      & 26.78      & 24.47      & 30.32      & 30.64      & 30.61         \\
                                  & Aug    & 25.08      & 26.67      & 24.26      & 30.08      & 30.5       & 30.4          \\
                                  & Sep    & 23.19      & 24.77      & 22.21      & 27.76      & 28.32      & 28.36         \\
                                  & Oct    & 19.43      & 20.96      & 18.88      & 23.97      & 24.56      & 24.65         \\
                                  & Nov    & 13.87      & 15.51      & 13.6       & 18.5       & 19.21      & 18.99         \\
                                  & Dec    & 9.48       & 11.09      & 9.23       & 13.9       & 14.84      & 14.35         \\
                                  & Annual & 17.42      & 19.1       & 16.94      & 22.35      & 22.94      & 22.84         \\
                                  \hline
\multirow{14}{*}{WS-10m (m/s)}    & Month  & Site No. 1 & Site No. 2 & Site No. 3 & Site No. 4 & Site No. 5 & Site No. 6\&7 \\
\hline
                                  & Jan    & 2.42       & 2.53       & 2.82       & 3.35       & 3.07       & 2.66          \\
                                  & Feb    & 2.7        & 2.67       & 3.11       & 3.62       & 3.26       & 2.93          \\
                                  & Mar    & 2.87       & 2.93       & 3.23       & 3.83       & 3.39       & 3.12          \\
                                  & Apr    & 2.92       & 3.05       & 3.22       & 3.89       & 3.41       & 3.18          \\
                                  & May    & 2.99       & 3.17       & 3.29       & 4.14       & 3.51       & 3.25          \\
                                  & Jun    & 3.16       & 3.4        & 3.54       & 4.83       & 3.96       & 3.55          \\
                                  & Jul    & 2.8        & 3.15       & 3.27       & 4.27       & 3.68       & 3.22          \\
                                  & Aug    & 2.68       & 3.09       & 3.16       & 4.3        & 3.73       & 3.23          \\
                                  & Sep    & 2.6        & 3.08       & 2.98       & 4.27       & 3.57       & 3.07          \\
                                  & Oct    & 2.24       & 2.79       & 2.53       & 3.38       & 3          & 2.58          \\
                                  & Nov    & 2.19       & 2.64       & 2.51       & 3.17       & 2.87       & 2.47          \\
                                  & Dec    & 2.24       & 2.52       & 2.59       & 3.21       & 2.95       & 2.52          \\
                                  & Annual & 2.65       & 2.92       & 3.02       & 3.86       & 3.37       & 2.98          \\
                                  \hline
\multirow{14}{*}{WD-10m (degree)} & Month  & Site No. 1 & Site No. 2 & Site No. 3 & Site No. 4 & Site No. 5 & Site No. 6\&7 \\
\hline
                                  & Jan    & 283.84     & 90.16      & 279.66     & 322.62     & 314.13     & 309.71        \\
                                  & Feb    & 285.22     & 196.79     & 283.7      & 322.34     & 316.15     & 294.94        \\
                                  & Mar    & 293.69     & 223.09     & 294.49     & 327.41     & 325.53     & 293.78        \\
                                  & Apr    & 304.77     & 263.96     & 308.09     & 271.54     & 327.28     & 270.34        \\
                                  & May    & 320.61     & 247.03     & 321.56     & 272.53     & 336.81     & 277.83        \\
                                  & Jun    & 263.39     & 118.94     & 321.25     & 342.26     & 335.66     & 338.99        \\
                                  & Jul    & 330.43     & 239.13     & 343.54     & 343.2      & 330.84     & 333.95        \\
                                  & Aug    & 299.83     & 132.62     & 348.05     & 342.36     & 329.27     & 334.11        \\
                                  & Sep    & 141.83     & 36.03      & 199.03     & 334.11     & 332.95     & 315.34        \\
                                  & Oct    & 214.16     & 43.7       & 229.72     & 191.15     & 336.41     & 169.64        \\
                                  & Nov    & 173.84     & 56.23      & 191.05     & 287.6      & 330.63     & 166.95        \\
                                  & Dec    & 229        & 63.81      & 256.65     & 324.68     & 319.47     & 226.89        \\
                                  & Annual & 261.72     & 142.62     & 281.4      & 306.82     & 327.93     & 277.71        \\
                                  \hline
\multirow{14}{*}{RH-2m (\%)}      & Month  & Site No. 1 & Site No. 2 & Site No. 3 & Site No. 4 & Site No. 5 & Site No. 6\&7 \\
\hline
                                  & Jan    & 46.33      & 47.37      & 52.29      & 47.33      & 44.35      & 42.62         \\
                                  & Feb    & 41.81      & 42.24      & 46.69      & 39.81      & 38.09      & 35.8          \\
                                  & Mar    & 36.64      & 36.18      & 40.82      & 33.79      & 33.09      & 30.26         \\
                                  & Apr    & 28.59      & 28.22      & 32.22      & 26.71      & 26.8       & 23.99         \\
                                  & May    & 27.73      & 27.06      & 31.37      & 24.7       & 24.83      & 22.5          \\
                                  & Jun    & 28.71      & 28.02      & 33.47      & 25.36      & 25.07      & 23.15         \\
                                  & Jul    & 30.97      & 30.63      & 35.83      & 27.98      & 27.89      & 25.77         \\
                                  & Aug    & 33.9       & 33.54      & 39.87      & 30.35      & 29.92      & 27.74         \\
                                  & Sep    & 37.5       & 37.33      & 45.43      & 34.37      & 33.35      & 30.57         \\
                                  & Oct    & 40.88      & 40.93      & 47.77      & 40.1       & 39         & 35.16         \\
                                  & Nov    & 44.3       & 44.34      & 49.82      & 45.56      & 43.65      & 40.45         \\
                                  & Dec    & 44.95      & 46.01      & 51.29      & 49.02      & 45.41      & 43.73         \\
                                  & Annual & 36.86      & 36.82      & 42.24      & 35.42      & 34.29      & 31.81         \\
                                  \hline
\multirow{14}{*}{PWV (mm)}        & Month  & Site No. 1 & Site No. 2 & Site No. 3 & Site No. 4 & Site No. 5 & Site No. 6\&7 \\
\hline
                                  & Jan    & 5.4        & 6.48       & 5.96       & 7.87       & 8.23       & 7.37          \\
                                  & Feb    & 5.59       & 6.58       & 6.12       & 7.85       & 8.18       & 7.36          \\
                                  & Mar    & 6.48       & 7.31       & 6.98       & 8.69       & 9.1        & 8.21          \\
                                  & Apr    & 7.72       & 8.47       & 8.21       & 9.96       & 10.48      & 9.48          \\
                                  & May    & 9.69       & 10.42      & 10.22      & 12.36      & 12.95      & 11.83         \\
                                  & Jun    & 8.96       & 9.68       & 9.46       & 11.41      & 11.9       & 10.8          \\
                                  & Jul    & 10.5       & 11.39      & 10.95      & 13.54      & 14.54      & 13.1          \\
                                  & Aug    & 11.69      & 12.61      & 12.2       & 14.72      & 15.92      & 14.31         \\
                                  & Sep    & 10.92      & 11.83      & 11.6       & 13.65      & 14.74      & 13.17         \\
                                  & Oct    & 10.45      & 11.39      & 11.22      & 13.76      & 14.84      & 13.1          \\
                                  & Nov    & 7.85       & 8.99       & 8.54       & 10.76      & 11.42      & 10.14         \\
                                  & Dec    & 6.06       & 7.18       & 6.68       & 8.77       & 9.13       & 8.18          \\
                                  & Annual & 8.44       & 9.36       & 9.01       & 11.11      & 11.79      & 10.59         \\
                                  \hline
\multirow{14}{*}{TCC (fractions)} & Month  & Site No. 1 & Site No. 2 & Site No. 3 & Site No. 4 & Site No. 5 & Site No. 6\&7 \\
\hline
                                  & Jan    & 0.28       & 0.25       & 0.32       & 0.21       & 0.19       & 0.2           \\
                                  & Feb    & 0.24       & 0.22       & 0.27       & 0.19       & 0.18       & 0.18          \\
                                  & Mar    & 0.21       & 0.2        & 0.24       & 0.18       & 0.17       & 0.17          \\
                                  & Apr    & 0.19       & 0.17       & 0.2        & 0.17       & 0.16       & 0.16          \\
                                  & May    & 0.17       & 0.15       & 0.17       & 0.15       & 0.15       & 0.15          \\
                                  & Jun    & 0.02       & 0.02       & 0.02       & 0.02       & 0.02       & 0.02          \\
                                  & Jul    & 0.01       & 0.01       & 0.02       & 0.01       & 0.02       & 0.02          \\
                                  & Aug    & 0.02       & 0.01       & 0.03       & 0.01       & 0.01       & 0.01          \\
                                  & Sep    & 0.04       & 0.02       & 0.08       & 0.02       & 0.02       & 0.02          \\
                                  & Oct    & 0.13       & 0.09       & 0.16       & 0.08       & 0.09       & 0.08          \\
                                  & Nov    & 0.22       & 0.19       & 0.26       & 0.18       & 0.17       & 0.17          \\
                                  & Dec    & 0.28       & 0.26       & 0.32       & 0.24       & 0.22       & 0.23          \\
                                  & Annual & 0.15       & 0.13       & 0.17       & 0.12       & 0.12       & 0.12  \\
\hline
\end{tabular}
}
\end{center}
      \end{table*}%
{AOD is the measurements of the total aerosols distributed through a column of air extending from the Earth’s surface to the top of the atmosphere. Usually numerical models and \textit{in situ} observations use mass concentration as the primary measure of aerosol loading, whereas most remote sensing methods retrieve AOD \citep{2008MNRAS.391..507V,chin2009atmospheric}.}\\
\begin{table}
\caption{Number of clear nights for each site during 2019.}
\begin{center}
\label{tab:phot_nights}
\begin{tabular}{lcc}
\hline
ID   & No. of Clear Nights & Ratio \\
\hline
1    & 289          & 79.18\%               \\
2    & 302          & 82.74\%               \\
3    & 276          & 75.62\%               \\
4    & 294          & 80.55\%               \\
5    & 290          & 79.45\%               \\
6\&7 & 290          & 79.45\%\\
\hline
\end{tabular}
\end{center}
\end{table}
{Due to the lack of \textit{in situ} observations, the remote sensing technique has been used in this work to estimate the AOD values over the candidate sites. The AOD data were collected from the Modern-Era Retrospective analysis for Research and Applications version 2 (MERRA-2) (\url{https://disc.gsfc.nasa.gov/datasets/M2TMNXAER_5.12.4/summary}).}

{These data are monthly averaged with spatial resolution of 0.5$^\circ$ X 0.625$^\circ$ at the visual (550 nm) wavelength during Oct-2018 to Oct-2020. Figure \ref{fig:AOD_Map} displays the average distribution of the AOD over Egypt including our candidate sites marked with black pins. The Figure in general clearly shows that the proposed sites have good AOD measurements in comparison with the other parts of Egypt.}
\begin{figure}
  \includegraphics[angle=0,width=0.45\textwidth]{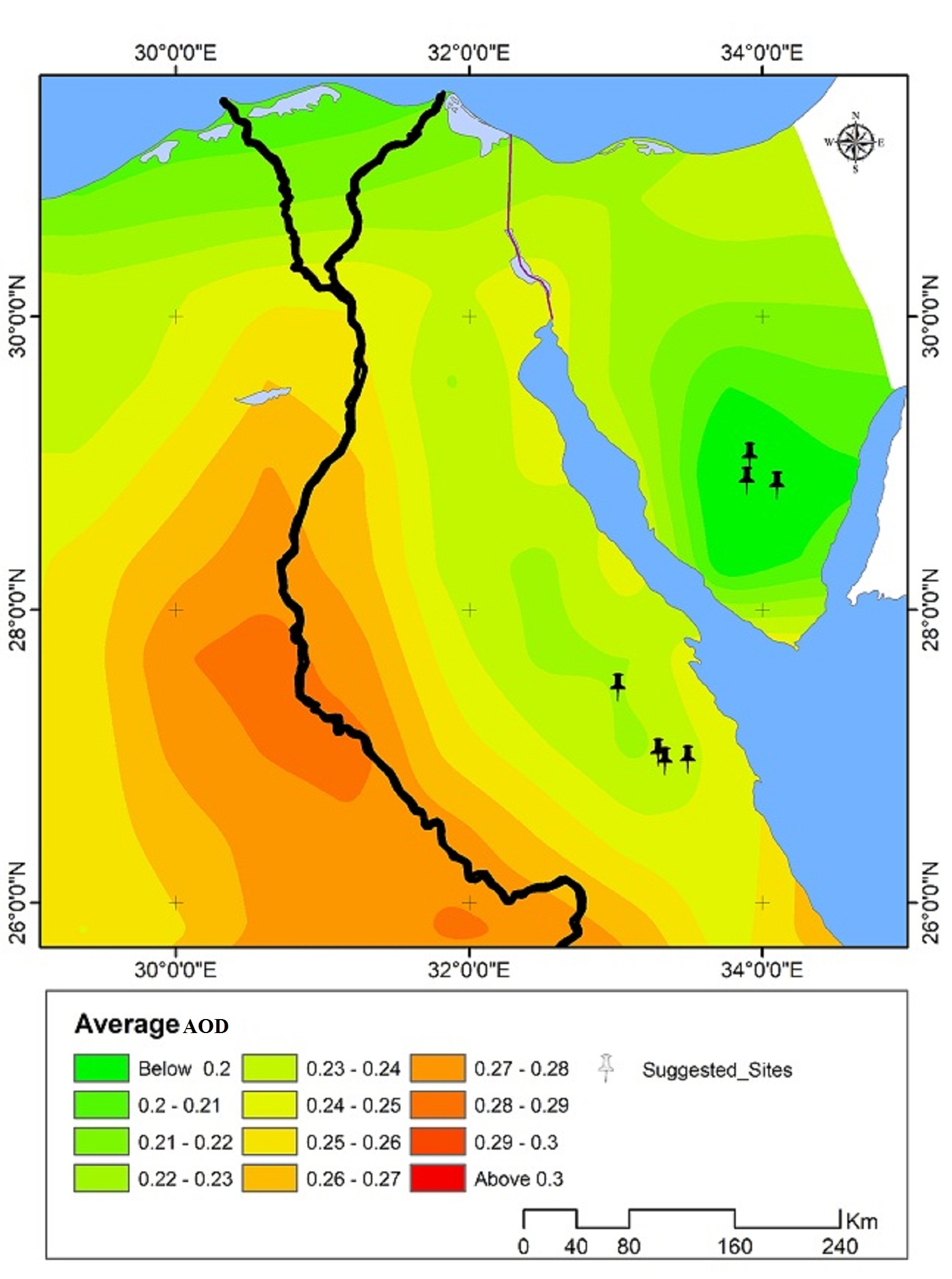}
  \caption{\label{fig:AOD_Map}Map of average AOD measurements over Egypt during the period Oct 2018 – October 2020, the black pins refer to our candidate sites.}
  \end{figure} %

{More detailed analysis has been performed on the candidate sites and the result is shown in Figure \ref{fig:Sites_AOD}. We noticed that, of the seven sites, sites No.2 and 3 have the lowest AOD values with an average of 0.181 and 0.184, respectively. Sites No.1 and 4 come next with an average AOD of 0.186 and 0.223, respectively.
Figure \ref{fig:Sites_AOD} shows that, the peak values occur during spring, which could be influenced by prevailing winds, notably the Elkhamaseen wind, while the lowest values are recorded during the winter. In addition to the wind blown, this pattern can be attributed to the effect of temperature on aerosol behavior: warmer air causes aerosols to ascend, while cooler air leads them to descend (see., \citet{alam2015particulate,kohil2017study,elshora2023evaluation}) The notion of temperature's impact on aerosol radiative effects finds substantiation in the work of \cite{goldstein2009biogenic}, although this insight coexists with conflicting findings presented by \cite{li2023spatial}. Therefore, any noticeable link between air temperature and AOD is probably part of a complex system of interactions and needs further investigation.

\begin{figure*}
  \includegraphics[angle=0,width=16.5cm]{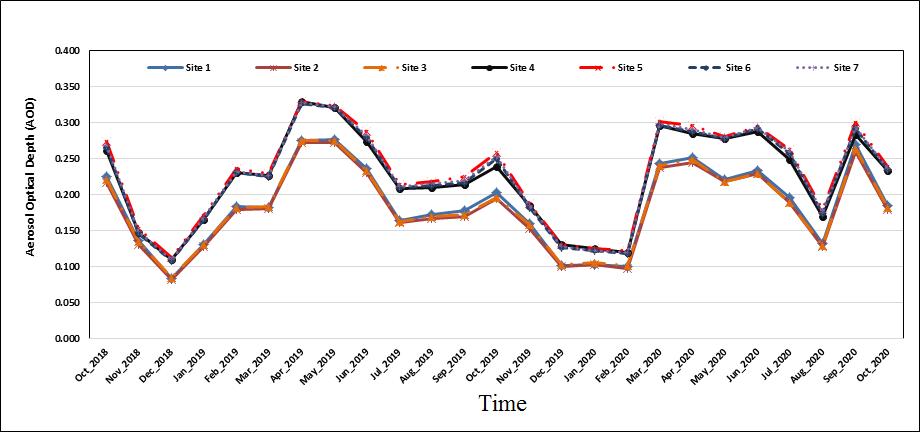}
  \caption{\label{fig:Sites_AOD}Average AOD values of the candidate sites during the period Oct 2018 – October 2020 at $\lambda $= 500 nm.}
  \end{figure*} %

 \subsection{Night Sky Brightness}
The most helpful way to estimate the impact of light pollution is to measure the Night Sky Brightness (NSB), from the ground location for the candidate site. There are various ways and devices to measure the NSB. One of the most widely and easiest devices used to measure the brightness of the sky is the Sky Quality Meter (SQM). Due to the lack of on site monitoring of the NSB, we relay on the zenith sky brightness \textbf{from} the 2015 world atlas of VIIRS measured in $\mathrm{mag./arcsec^2}$.
The results of the seven candidate sites are listed in Table \ref{tab:NBS}.
\begin{table}
	\caption{NSB measurements over our candidate sites as extracted from VIIRS 2015 world atlas.}
	\label{tab:NBS}
		\renewcommand{\arraystretch}{1.1}
		\begin{tabular}{|c|c|c||c|c|}
			\hline
             Site-No        &NSB & \\
                        &             ($mag./arcsec^2$)&  \\
			\hline
			
			1 & 21.97    &   \\
			2 & 21.97  &   \\
			3 & 21.98   & \\
			4 & 21.96 &  \\
			5& 21.89  &  \\
			6& 21.94 &  \\
            7& 21.93 &  \\
			\hline
		\end{tabular}%
	
\end{table} %
Table \ref{tab:NBS} indicate that the NSB value at all the candidate sites is good, however the first four sites are slightly darker than the other sites.
\section{Seismic Hazards}\label{sec:s_hazard}
{For the safe design and operation of the observatory, it is preferable to select a site with low seismic hazard, characterized by low seismicity, low ground shaking intensity, and a safe distance from faults. Therefore, a comprehensive understanding and assessment of seismic input or seismic hazard, including ground motion peak parameters and response spectra (RS), are crucial for the selection of suitable sites in a multi-parameter study. If multiple sites have similar observational, astronomical, and accessibility qualities, they may differ in terms of the severity of ground shaking intensities, such as Peak Ground Acceleration  (PGA) or maximum acceleration and RS, which require varying levels of seismic fortification and incur higher costs.}

{Engineers also require seismic input to construct structures with seismic resistance that ensure good performance and protection during earthquakes, minimizing risks and safeguarding scientists and potentially expensive instruments and technologies. Simultaneously, it is important to ensure the functionality of the observatory during earthquake events. Therefore, a reliable seismic hazard assessment (SHA) is essential, employing physics-based multi-scenario seismic hazard analysis to mitigate future losses. The candidate sites are located in a region known for relatively high seismic hazard, including the north Red Sea and the gulfs of Suez and Aqaba \citep{mohamed2012seismic,sawires2016updated,hassan2017insight,hassan2017update}. The seismicity map for the proposed sites is shown in Figure \ref{sismis-map}.
\begin{figure*}
  \centering
\includegraphics[width=12cm,height=12cm]{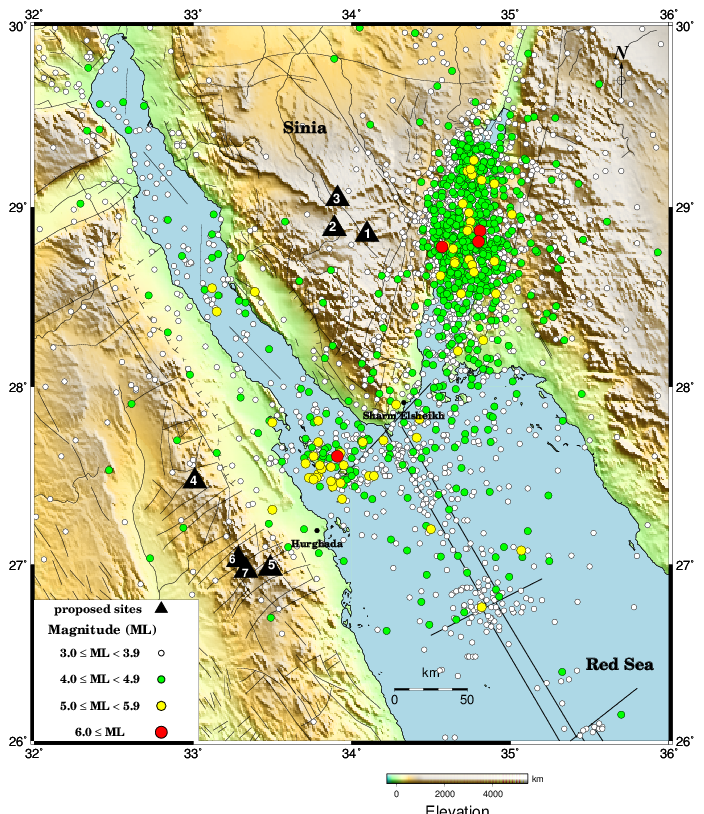}
\caption{Location of the proposed sites (black triangle) combined with seismicity (till 2020) and surface faults in the region.}\label{sismis-map}
\end{figure*}
\begin{figure*}
  \centering
\includegraphics[width=15cm,height=9cm]{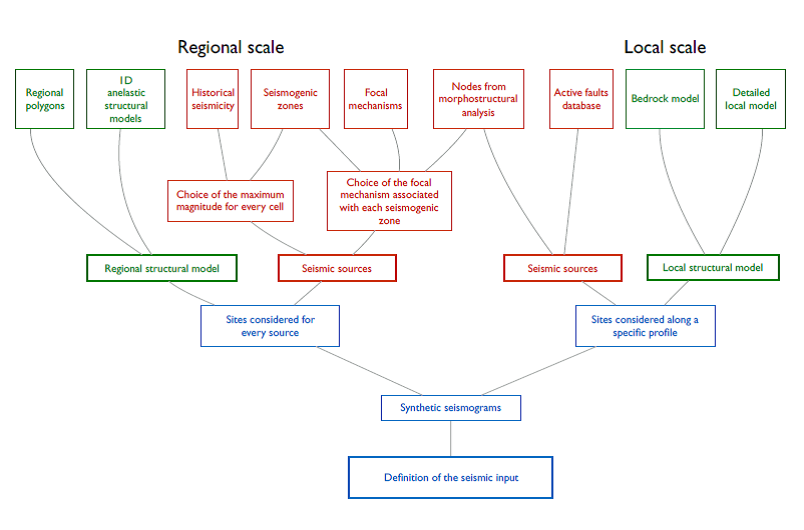}
\vspace*{8pt}
\caption{Flow chart of the different steps in the NDSHA approach for the regional-scale analysis.}\label{block-diagramz}
\end{figure*}
%

During the last decades, significant earthquakes have hit the Gulf of Aqaba, Gulf of Suez and northern Red Sea. Seismicity and seismotectonic setting of these sources have been studied by many \citep[e.g.][]{ali2019present,badawy2020source,badreldin2022dynamic} in order to get a better understanding of the present-day stress regime for them.

In this study, the Neo-Deterministic Seismic Hazard Analysis approach (NDSHA) is utilized to provide the seismic input parameters required for the multi-parameter study in the site selection of a new Ground-Based telescope. A detailed description of the NDSHA methodology can be found in \cite{panza2001seismic} and its updates and validations in \cite{panza2012seismic}and \cite{magrin2016broadband}. The NDSHA approach enables the estimation of ground motion parameters, such as maximum acceleration, velocity, displacement, and RS, with a high degree of reliability. It utilizes information about potential seismic sources, seismicity, and the mechanical properties of the medium between these sources and the sites of interest (for further details, refer to \cite{panza2001seismic, panza2012seismic, bela2012seismic,magrin2016broadband,hassan2017insight,panza2020ndsha,kossobokov2022seismic}. NDSHA employs scenario-based methods for seismic hazard analysis, constructing realistic synthetic time series for earthquake scenarios. It is particularly suitable for computing ground motion parameters at 1 and 10 Hz cut-off frequencies for a set of 1D structural models at epicentral distances greater than the focal depth of the source, at different spatial scales (\ref{block-diagramz}).
Starting from the available knowledge about Earth's structure and the propagation of seismic waves, as well as information about seismic sources and seismicity in the study area, it is possible to realistically compute synthetic seismograms. These synthetic seismograms allow the quantification of relevant parameters such as ground motion acceleration, velocity, displacement, and other parameters important for seismic engineering (e.g., percentiles, resultant, maximum).

\subsection{Input Parameters for NDSHA Computation}%
In the current work, input data for seismic hazard assessment in the NDSHA framework are taken from \citet{hassan2017insight} and then updated. These data are then used to compute the seismic input (Maximum Acceleration)  at the sites of interest for the purpose of multi-parameter site selection for an astronomical observatory. The ground motion maps for Egypt were computed by \citet{hassan2017insight}, utilizing revised and up-to-date input data, including earthquake catalogue, seismotectonic zones with their representative focal mechanisms, and structural models (Figure \ref{sism_3}).
In the current work, we found that the earthquake catalogue is the only ingredient that should be upgraded since more data are available. The earthquake catalogue used by the NDSHA package requires the availability, as complete as possible, of earthquakes with M$\ge$ 5, which are capable of generating significant ground motion. The initial dataset is the earthquake catalogue of \cite{hassan2017insight}, updated until 2020, with a proper comparison with other available historical or instrumental earthquake catalogues from national (e.g., \url{ http://ensn.nriag.sci.eg/}) and international sources (e.g., European Mediterranean Seismological Center (EMSC) \url {http://www.emsc-csem.org/}; International Seismological Center (ISC) bulletins \url {http://www.isc.ac.uk/iscbulletin/search/}). All this information has been used to compile a uniform and, as much as possible, complete earthquake catalogue (smoothed seismicity is shown in Figure \ref{im-az}). The pre-instrumental earthquake catalogue is taken from the revised and quality-controlled catalogue of NRIAG. The instrumental earthquake catalogue is compiled using all available national (either published or unpublished) and international catalogues, as well as existing publications about seismicity and source mechanisms. The earthquake's magnitude is converted into the moment magnitude (Mw) scale for homogenization purposes (Figure \ref{im-az}).
Twenty zones have been defined in \cite{hassan2017insight} based on the available information, such as earthquake catalogues, refined focal mechanism solutions, surface geology, geophysical studies, surface faults, GPS data, crustal structure, and other related studies. In this work, seismogenic zones model from the work of \cite{hassan2017insight} combined with seismogenic nodes (earthquake-prone areas) were obtained based on morphostructural analysis and pattern recognition techniques \citep{gorshkov2019seismogenic, gorshkov2022identifying} (Figure \ref{im-az} \& \ref{im-bz}).
According to the geologic maps available for the study area, the proposed sites are located in mountainous areas dominated by igneous and metamorphic rock. Therefore, a structural model at the rock site is sufficient for hazard computation. In this work, eight average anelastic structural polygons are delineated by \cite{hassan2017insight} based on all the available crustal structure data from seismic reflection and gravity surveys, as well as the velocity models adopted by the Egyptian National Seismological Network (ENSN) and NRIAG for earthquake locations in Egypt (Figure \ref{im-cz}).
\subsection{Seismic Hazard Input Parameters} \label{accela}%
The synthetic seismograms have been computed at a 10 Hz cut-off frequency, and the seismic sources within the seismogenic zones are treated with proper seismic source approximation \citep{parvez2011long}, combining the input parameters described above. The maximum acceleration or 95$^{th}$ percentile (hereinafter referred to as A) values for the seven sites were computed for the vertical and horizontal components. Then, the maximum value at each site is extracted and plotted in Figure \ref{sismis-4}. The map of maximum acceleration indicates that sites 1, 2, and 5 are exposed to high seismic hazards relative to the others (Red circle in Figure \ref{sismis-4}). Sites 6 and 7 are subjected to moderate seismic hazard (Orange circle in Figure \ref{sismis-4}), while sites 3 and 4 are exposed to low seismic hazard (Yellow circle in Figure \ref{sismis-4}).
For earthquake engineering purposes, the Maximum Credible Seismic Input (MCSI) represents a reliable estimation of the expected ground shaking level for a specific site, independent of the occurrence of earthquakes that could affect the investigated area. In NDSHA, thousands of ground motion time histories needed for engineering analysis are simulated, and all of these parameters can be summarized in MCSI. It provides a reliable estimation of the upper-bound level of shaking that could occur at a selected site, neglecting the probability of occurrence. The aim is to define a reliable and effective design seismic input (\citet{fasan2017advanced, Rugarli_Vaccari_Panza_2019}).
Regarding the physical definition of the MCSI response spectrum, it is thoroughly described in \citet{fasan2017advanced} and \citet{Rugarli_Vaccari_Panza_2019}. According to this definition, for each seismogenic source, n-scenarios (in terms of magnitude, epicentral distance, and focal mechanism of the earthquake) have been considered, and the obtained spectral acceleration values (SA) are compared, selecting the maximum median.
In this study, MCSI is computed at bedrock-MCSIBD with a 10 Hz cut-off frequency for all the selected sites shown in Figure \ref{block-diagramz}. Due to the complexity of the rupture process on a fault (and the implicit impossibility of deterministically predicting future events), a hundred of its kinematic realizations have been used. The results, including the median and 95$^{th}$ percentile (median +2 $\sigma$) of the computed MCSI at the selected sites (sites 1 to 7), are shown in Figure \ref{sismis-6}. These curves represent the response spectra developed by combining the results of all individual scenarios within the area of influence, considering a hundred rupture realizations and 5\% damping.
The computed MCSI spectra at sites 1, 2, and 5 (Figure \ref{sismis-6} a, b, and e, respectively) show high hazard at short periods (0.1-1.0 s). Site 5 also shows another peak at longer periods (1.3-3.0 s). Sites 6 and 7 (Figure \ref{sismis-6} f and g, respectively) exhibit two peaks, both indicating moderate seismic hazard. One peak occurs at short periods (0.1-1.0 s), and the other occurs at longer periods (1.3-3.0 s). Sites 3 and 4 (Figure \ref{sismis-6} c and d, respectively) experience low hazard and show peaks in the period range of 0.1-1.0 s, while site 4 shows an additional peak in the period range of 1.3-3.0 s.
Considering the current results, the peak ground motion acceleration in Figure \ref{sismis-4}, and the response spectra curves for all sites shown in Figure \ref{sismis-4}, it can be concluded that site 3 poses the least seismic hazard. A summary of this section is presented in Table \ref{tab:sei_class}, where the sites are classified into three groups based on their expected hazard levels. This table can guide astronomers and decision-makers in making informed decisions about future observatory locations.
\begin{figure*}
\centering
\begin{subfigure}{.5\textwidth}
  \centering
  \includegraphics[width=8cm,height=7cm]{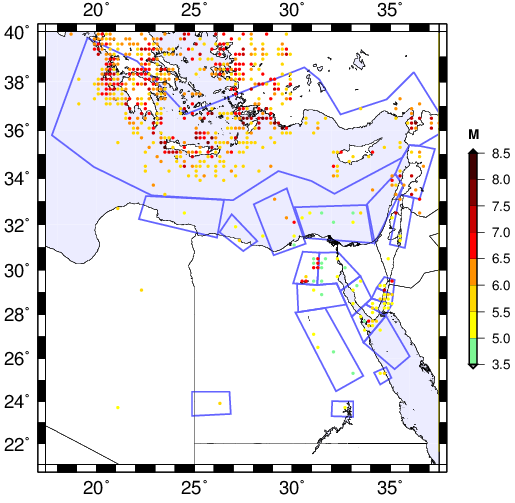}
  \caption{}
  \label{im-az}
\end{subfigure}%
\vspace*{8pt}
\begin{subfigure}{.5\textwidth}
  \centering
  \includegraphics[width=8cm,height=7cm]{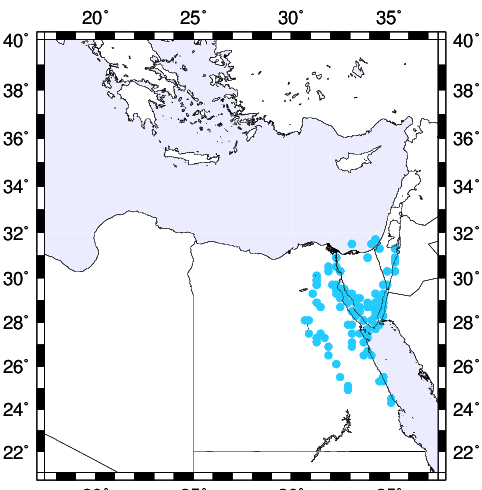}
    \caption{}
  \label{im-bz}
\end{subfigure}%
\begin{subfigure}{.5\textwidth}
  \centering
  \includegraphics[width=8cm,height=7cm]{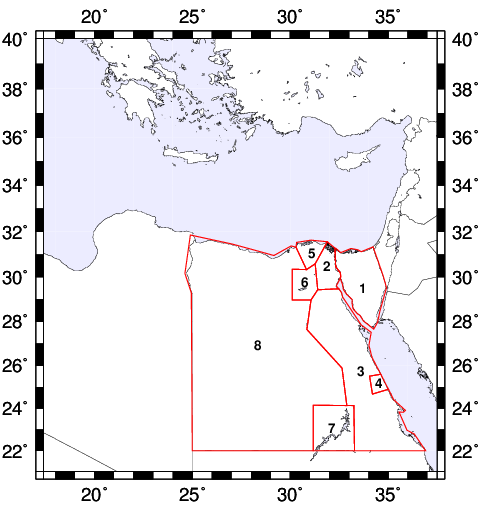}
  \caption{}
  \label{im-cz}
\end{subfigure}%
\hspace*{8pt}
\caption{(a) Earthquake catalogue Mw$\geq$5 till 2020 plotted with the seismogenic zones model; (b) Seismogenic nodes; (c) Crustal structure model
.}\label{sism_3}
\end{figure*}
\begin{figure*}
  \centering
\includegraphics[width=13cm,height=13cm]{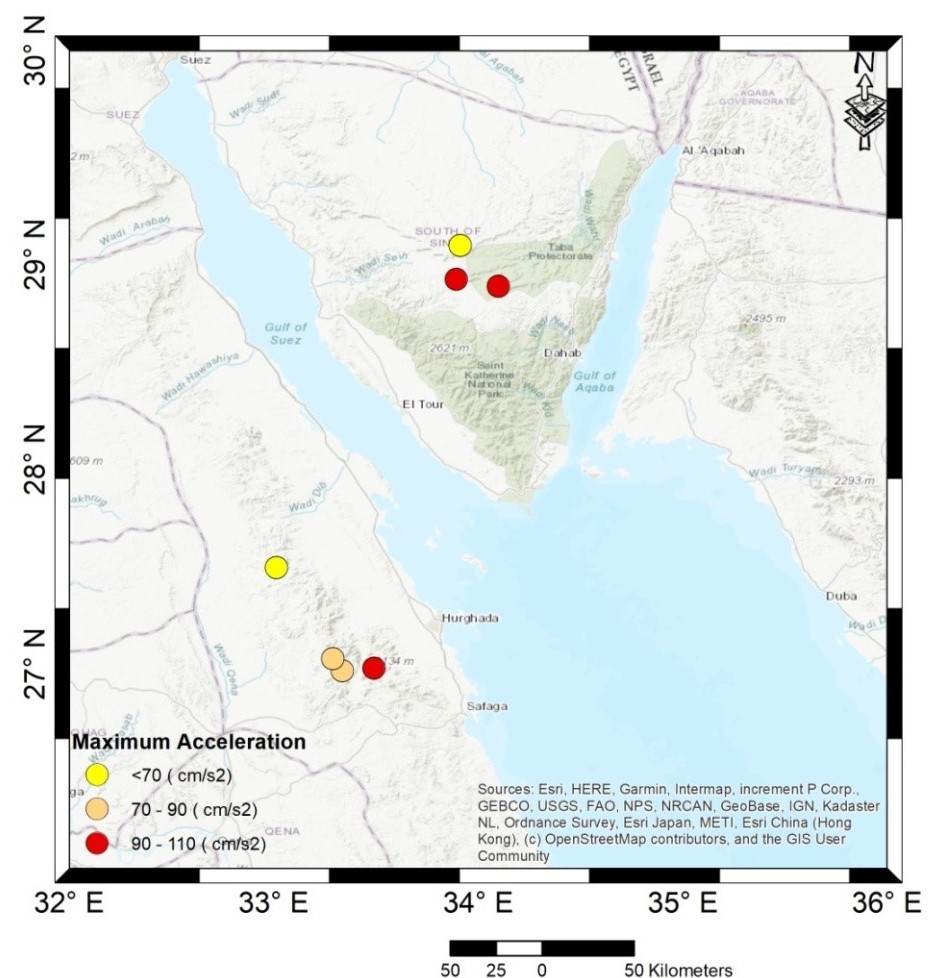}
\vspace*{8pt}
\caption{Peak ground motion acceleration map for the proposed sites. The red circles represent the highest seismic hazard sites, moderate seismic hazard are marked with orange circles while low seismic hazard sites are presented with yellow circles.}\label{sismis-4}
\end{figure*}
\begin{figure*}
  \centering
\includegraphics[width=14cm,height=14cm]{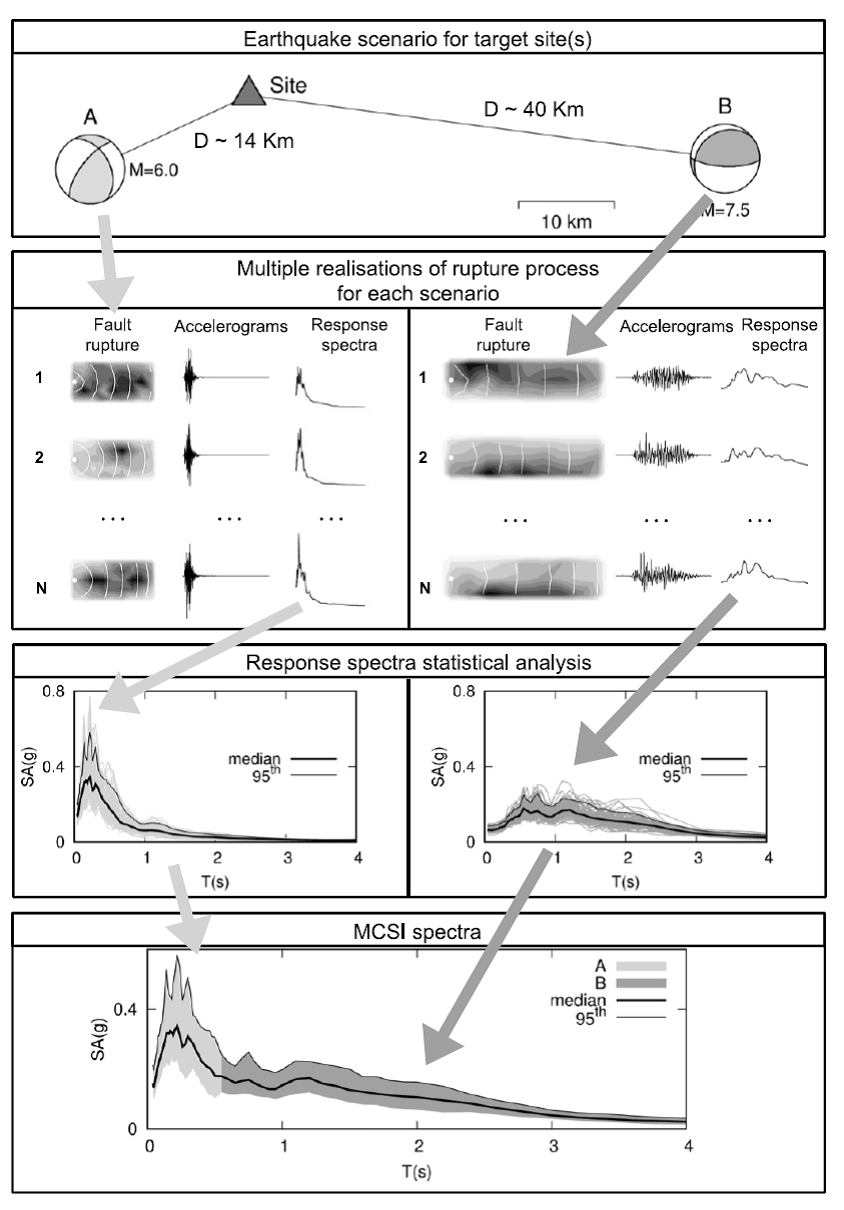}
\vspace*{8pt}
\caption{Description of the MCSI definition procedure.}\label{sismis-5}
\end{figure*}
\begin{figure*}
\centering
\begin{subfigure}{.5\textwidth}
  \centering
  \includegraphics[width=8cm,height=5.1cm]{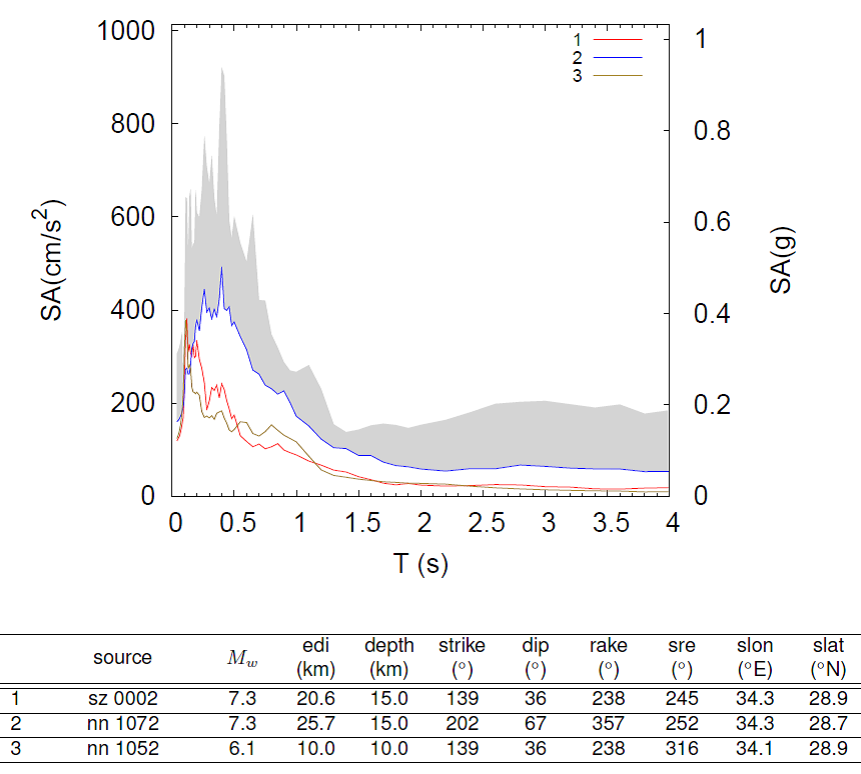}
  \caption{}
  \label{im-aaz}
\end{subfigure}%
\begin{subfigure}{.5\textwidth}
  \centering
  \includegraphics[width=8cm,height=5.1cm]{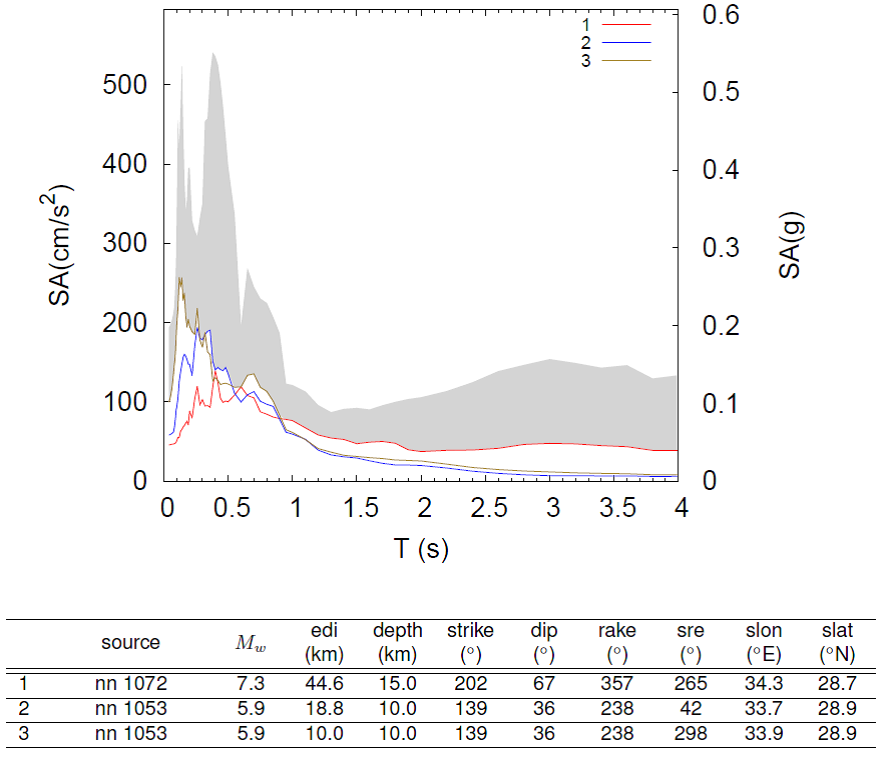}
    \caption{}
  \label{im-bbz}
\end{subfigure}%

\begin{subfigure}{.5\textwidth}
  \centering
  \includegraphics[width=8cm,height=5.1cm]{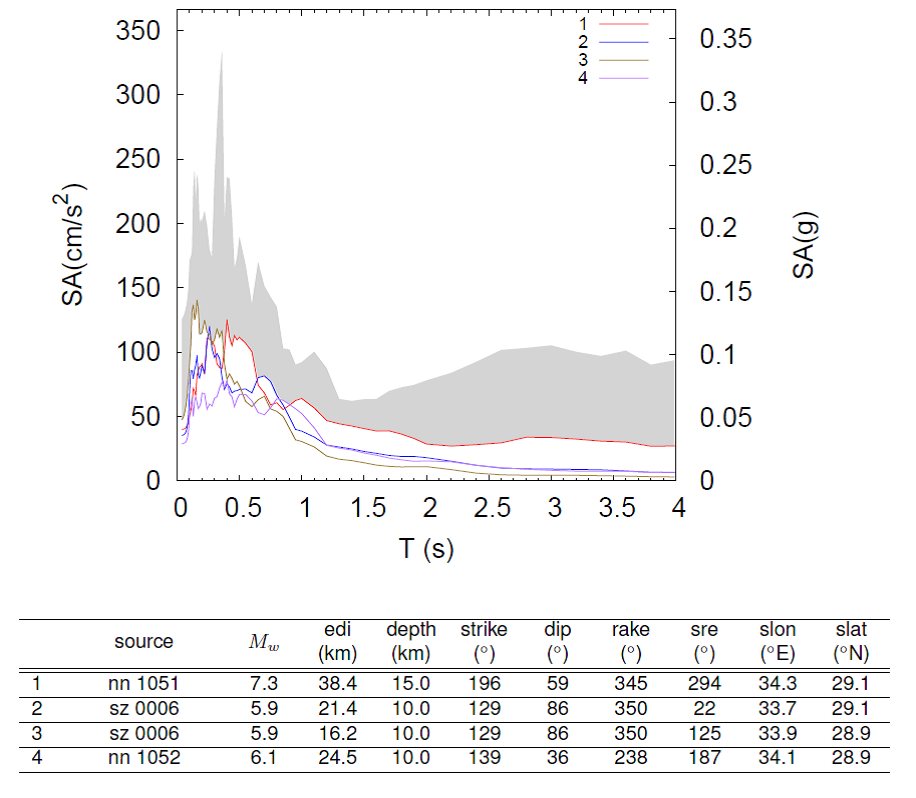}
  \caption{}
  \label{im-ccz}
\end{subfigure}%
\begin{subfigure}{.5\textwidth}
  \centering
  \includegraphics[width=8cm,height=5.1cm]{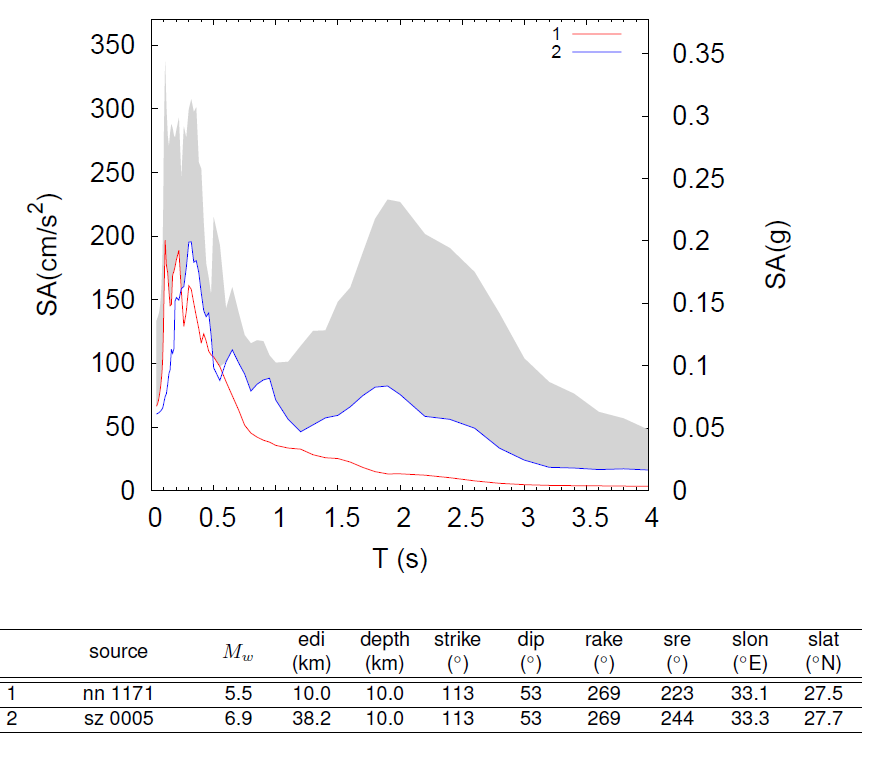}
  \caption{}
  \label{im-dz}
\end{subfigure}%

\begin{subfigure}{.5\textwidth}
  \centering
  \includegraphics[width=8cm,height=5.1cm]{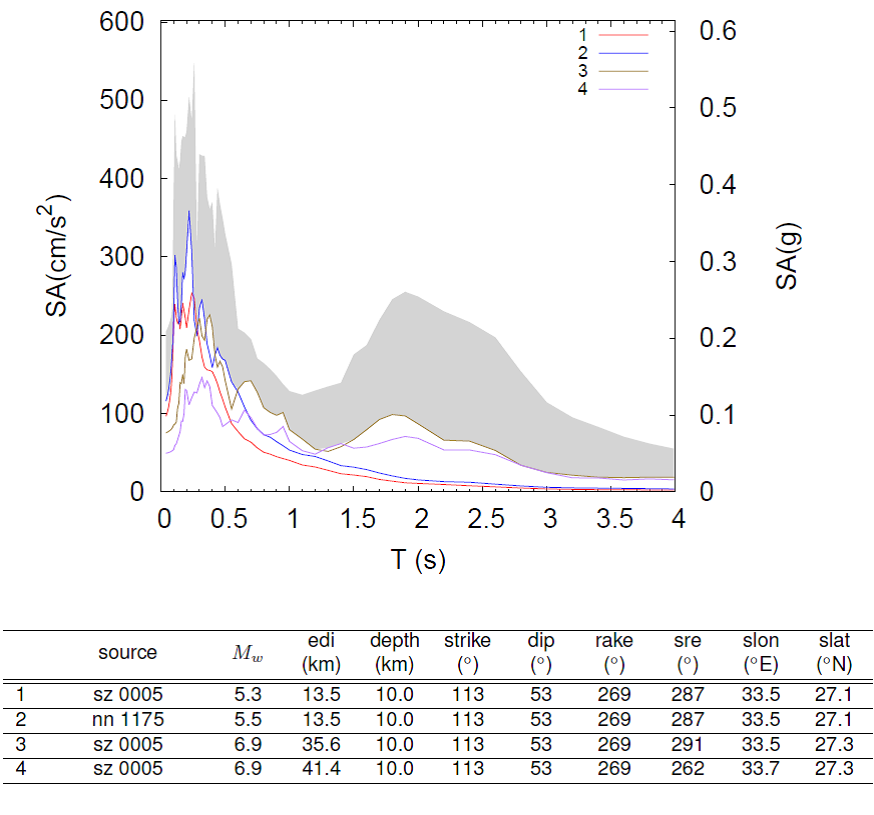}
  \caption{}
  \label{im-ez}
\end{subfigure}%
\begin{subfigure}{.5\textwidth}
  \centering
  \includegraphics[width=8cm,height=5.1cm]{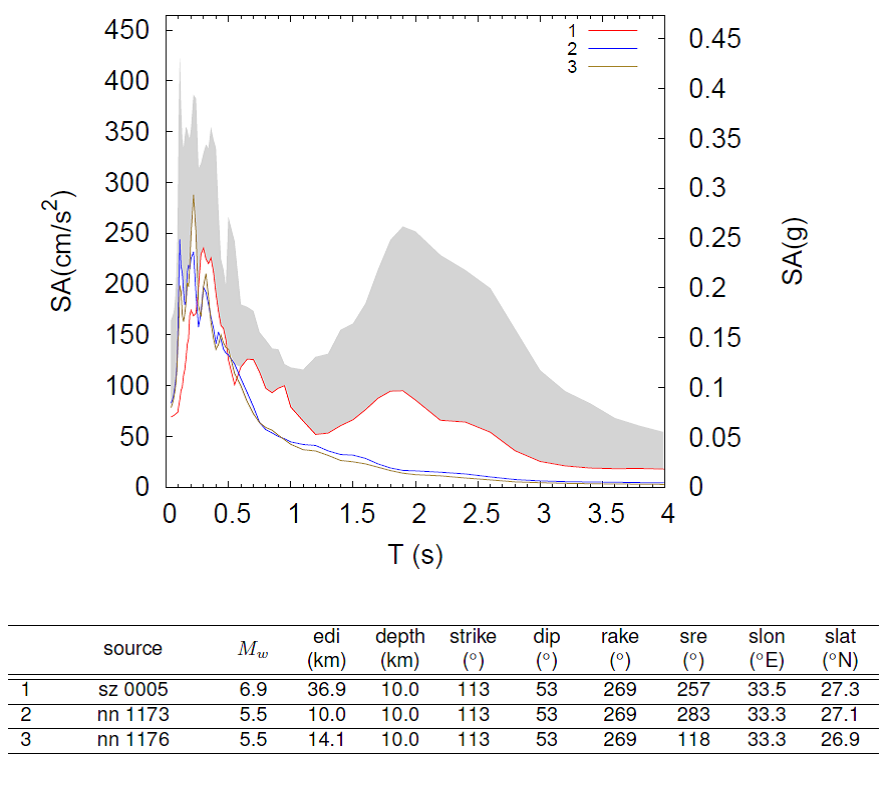}
  \caption{}
    \label{im-fz}
\end{subfigure}%

\begin{subfigure}{.5\textwidth}
  \centering
  \includegraphics[width=9cm,height=5.1cm]{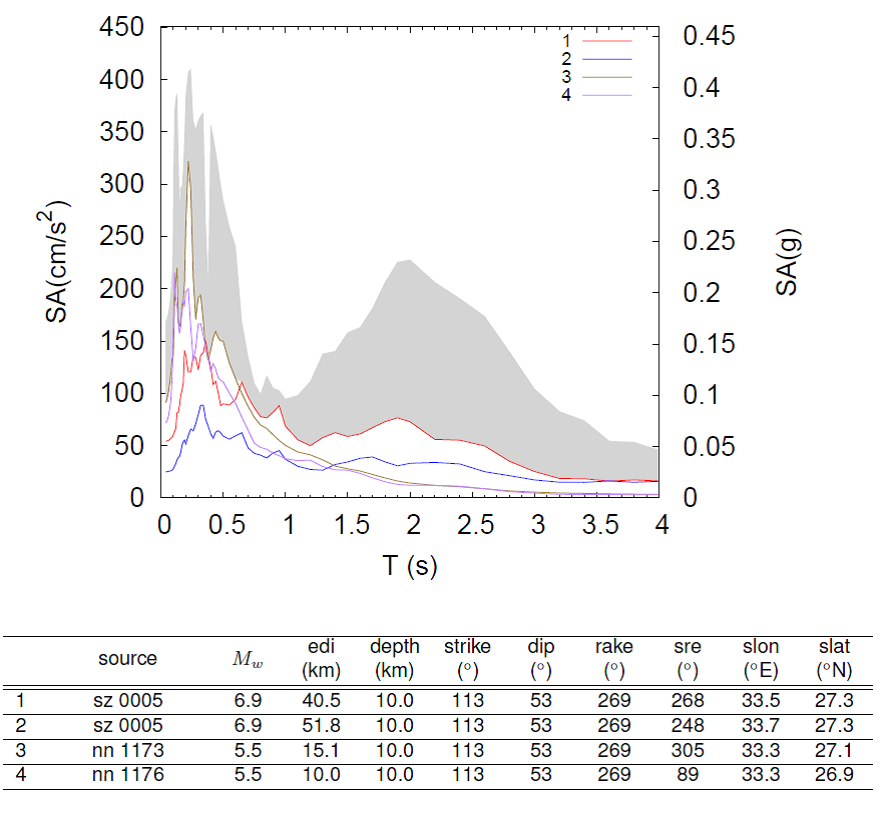}
  \caption{}
  \label{im-gz}
\end{subfigure}%
\caption{a) The MCSIBD is set equal to the value of the 50-95$^{th}$ percentile(shaded grey zone) for proposed sites 1 (a), 2 (b), 3 (c), 4 (d), 5 (e), 6 (f), 7 (g) as shown in Figure \ref{sismis-map}.}\label{sismis-6}
 \end{figure*}
\begin{table}
\caption{Seismic classification of the candidate sites.}\label{tab:sei_class}
\begin{tabular}{lclclc}
\hline
Site No. & Seismic hazard \\
\hline
3 and 4 & Low \\
6 and 7 & Moderate \\
1, 2 and 5 & High \\
\hline
\end{tabular}
\end{table}
\section{Conclusions and Recommendations}\label{sec:conc_reco}
{After combining all the results from different parameters, we found that of the 7 candidate sites only four sites can satisfy the initial conditions for building a future Egyptian observatory (see, Table \ref{tab:overallscore-results}). Nevertheless, we calculated the overall score of the estimated parameters over the four sites in order to evaluate them.}

{Table \ref{tab:overallscore-results} summarizes the overall score of the best four selected sites. The values of all  parameters involved are presented in the previous tables, except the wind direction (WD-10m). To consider the wind direction, we estimated the variance of its values across the year and presented the results in column five (Var (WD-10m)), while the letter A (column 10) represents the ground motion acceleration (see, sec \ref{accela}). The scores are computed by averaging the normalized values of all parameters, as in \citep{41,55,1,11,39,27,44,45}. In particular, we can express the computation of the overall score as
\begin{align}
    \text{Score}_i = \frac{1}{N_\text{S}}\left [ \frac{\text{AOD}_i^*}{\max \left (\text{AOD}_i^*\right)}+\frac{\text{NSB}_i}{\max \left (\text{NSB}_i\right)}+ \frac{\text{Ws}_i^*}{\max \left (\text{Ws}_i^*\right)} \right.  \nonumber \\
      \left.+ \frac{\text{Var}_i^*}{\max \left (\text{Var}_i^*\right)}+ \frac{\text{RH}_i^*}{\max \left (\text{RH}_i^*\right)} + \frac{\text{PWV}_i^*}{\max \left (\text{PWV}_i^*\right)} \right. \nonumber \\
      \left. + \frac{\text{TCC}_i^*}{\max \left (\text{TCC}_i^*\right)} + \frac{\text{AT}_i^*}{\max \left (\text{AT}_i^*\right)} + \frac{\text{A}_i^*}{\max \left (\text{A}_i^*\right)} \right. \nonumber \\
      \left. + \frac{\text{Clear Nights}_i}{\max \left (\text{Clear Nights}_i\right)} \right],
\end{align}
\begin{align}
   & \text{AOD}_i^* = \frac{1}{\text{AOD}_i},\;
    \text{Ws}_i^* = \frac{1}{\text{Ws}_i},\;
    \text{Var}_i^* = \frac{1}{\text{Var}_i},\;
    \text{RH}_i^* = \frac{1}{\text{RH}_i}, \nonumber \\
   & \text{PWV}_i^* = \frac{1}{\text{PWV}_i},\;
    \text{TCC}_i^* = \frac{1}{\text{TCC}_i}, \;
    \text{AT}_i^* = \frac{1}{\text{AT}_i},\;
    \text{A}_i^* = \frac{1}{\text{A}_i}, \nonumber
\end{align}
where $i =1 \cdots N_\text{S}$, $N_\text{S}$ is the number of sites, and $\text{Score}_i$ indicates the overall score of each site.

The results clarify that site No. 3 has the best overall score of 0.9128 followed by site No. 4 with a score of 0.8912. Finally, site No. 7 and site No. 2 have almost the same scores of 0.8571 and 0.8563, respectively.}

{The results from this work provide an important component in the multi-parameters site selection analysis and can guide the decision makers on which site is preferable from meteorological, observational and earthquake hazard point of view.}

{Ultimately, on site observations for the meteorological parameters, seeing with different tools, long time light pollution observations and detailed seismic hazard analysis for the new sites are highly recommended for further details of astronomical sites selection as well as sites testing.}
\begin{table*}
    \centering
    \caption{Overall score summary for the four proposed sites.}\label{tab:overallscore-results}
   \resizebox{\textwidth}{0.12\textwidth}{
    \begin{tabular}{|l|l|l|l|l|l|l|l|l|l|l|l|l|l|}
    \hline
       Site No. & AOD & NSB  &  Ws & Var(Wd) & RH  & PWV & TCC & AT & A  & Clear Nights & Overall Score \\
     &    &  ($mag./arcsec^2$) & (arcsec) & m/s &  \% & (mm) & (fraction) & ºC  & (gal) &  & &\\
\hline
  3    & 0.184 & 21.98 & 3.02 & 2784.3451 & 42.24 & 9.01 & 0.17 & 16.94 & 50.08 & 276 & 0.9128 \\ \hline
  4    & 0.223 & 21.96 &3.86 & 1999.1101 & 35.42 & 11.11 & 0.12 & 22.35 & 61.52 & 294 & 0.8912 \\ \hline
  7    & 0.227 & 21.93 & 2.98 & 3607.1069 & 31.81 & 10.59 & 0.12 & 22.84 & 86.53 & 290 & 0.8571\\ \hline
  2    & 0.181 & 21.97 & 2.92 & 7507.4972 & 36.82 & 9.36 & 0.13 & 19.1 & 93.44 & 302 & 0.8563 \\ \hline
    \end{tabular}}
\end{table*}
%

\section*{acknowledgements}
The authors would like to express their appreciation to the anonymous reviewers for their valuable insights and suggestions, which have significantly elevated the quality of this paper. The expertise and considerate feedback provided by the reviewers have immensely contributed to the refinement of the ideas presented in this manuscript.\\
We warmly thank Profs. G. El-Qady (NRIAG's Director), M. Nouh, H. Selim, Y. Azzam, A. Shaker and our colleagues at the Department of Astronomy, NRIAG for their fruitful discussion and continued support through the work.
{M.Darwish  and S. Saad would like to acknowledge the Science and Technology Development Fund (STDF) N5217, Academy of Scientific Research and Technology (ASRT), Cairo, Egypt and Kottamia Center of Scientific Excellence for Astronomy and Space Sciences (KCSEASSc), National Research Institute of Astronomy and Geophysics.}
Hany M. Hassan and Hazem Badreldin are grateful to the Department of Mathematics and Geosciences (DMG) staff, at Trieste University, Italy. The seismic hazard computation part was carried out using DMG facilities.

%
\section*{Data Availability}
The data underlying this article will be shared on reasonable request to the corresponding author.
\bibliographystyle{mnras}
\bibliography{ref}
%
 \bsp	
\label{lastpage}}
   \end{document}